\documentclass[11pt,twoside]{./atmp}
\usepackage{amsmath, amsfonts, amssymb, amscd, enumerate, mathrsfs}
\usepackage[all]{xy}
\usepackage{color}
\theoremstyle{plain}
\newtheorem{theo}{Theorem}[section]
\newtheorem{prop}[theo]{Proposition}
\newtheorem{cor}[theo]{Corollary}
\theoremstyle{definition}
\newtheorem{rem}[theo]{Remark}
\newtheorem{example}[theo]{Example}
\newtheorem{definition}[theo]{Definition}

\newenvironment{pf}{\noindent{\it Proof. }}{$\square$\par\medskip}
\newenvironment{pfns}{\noindent{\it Proof. }}{\par\medskip}
\renewcommand{\=}{\overset{\operatorname{def}}{=}}
\newcommand{\beq}{\begin{equation}}
\newcommand{\eeq}{\end{equation}}
\renewcommand{\a}{\alpha}
\renewcommand{\b}{\beta}

\renewcommand{\d}{\delta}
\newcommand{\e}{\epsilon}
\newcommand{\ve}{\varepsilon}

\newcommand{\g}{\gamma}
\newcommand{\h}{\eta}

\renewcommand{\l}{\lambda}

\renewcommand{\o}{\omega}

\newcommand{\s}{\sigma}

\newcommand{\z}{\zeta}

\newcommand{\G}{\Gamma}

\renewcommand{\L}{\Lambda}
\renewcommand{\O}{\Omega}

\newcommand{\bA}{\mathbb{A}}
\newcommand{\bB}{\mathbb{B}}
\newcommand{\bC}{\mathbb{C}}
\newcommand{\bD}{\mathbb{D}}
\newcommand{\bR}{\mathbb{R}}
\newcommand{\bZ}{\mathbb{Z}}

\newcommand{\bH}{\mathbb{H}}
\newcommand{\bK}{\mathbb{K}}

\renewcommand{\gg}{\mathfrak{g}}
\newcommand{\gh}{\mathfrak{h}}
\newcommand{\gk}{\mathfrak{k}}

\newcommand{\go}{\mathfrak{o}}
\newcommand{\gp}{\mathfrak{p}}

\newcommand{\gX}{\mathfrak{X}}

\newcommand{\so}{\mathfrak{so}}

\newcommand{\ggl}{\mathfrak{gl}}
\newcommand\GL{\mathrm{GL}}

\newcommand\Spin{\mathrm{Spin}}

\newcommand{\cA}{\mathcal{A}}

\newcommand{\cC}{\mathcal{C}}
\newcommand{\cD}{\mathcal{D}}
\newcommand{\cE}{\mathcal{E}}

\newcommand{\cG}{\mathcal{G}}
\newcommand{\cH}{\mathcal{H}}

\newcommand{\cL}{\mathcal{L}}

\newcommand{\cR}{\mathcal{R}}
\newcommand{\cS}{\mathcal{S}}
\newcommand{\cT}{\mathcal{T}}
\newcommand{\cU}{\mathcal{U}}

\newcommand\GamD[3]{\cH^{\ #1}_{#2#3}}
\newcommand\Gamd[3]{\bA{}^{\ #1}_{#2#3}}
\newcommand\Gamb[3]{\bB{}^{\ #1}_{#2#3}}
\newcommand\Gamc[4]{\bC{}^{\ \ #1}_{#2#3#4}}

\renewcommand{\square}{\kern1pt\vbox
{\hrule height 0.6pt\hbox{\vrule width 0.6pt\hskip 3pt
\vbox{\vskip 6pt}\hskip 3pt\vrule width 0.6pt}\hrule height0.6pt}\kern1pt}

\DeclareMathOperator\Id{Id}

\DeclareMathOperator{\Span}{Span}

\DeclareMathOperator\End{End\;}

\DeclareMathOperator\ad{ad}

\renewcommand\Re{\operatorname{Re}}
\renewcommand\Im{\operatorname{Im}}

\newcommand{\Hom}{{\operatorname{Hom}}}
\newcommand{\Sym}{{\operatorname{Sym}^g}}

\newcommand{\wt}{\widetilde}
\newcommand{\wh}{\widehat}

\newcommand{\n}{\nabla}

\newcommand{\be}{\begin{equation}}
\newcommand{\ee}{\end{equation}}

\def\<#1,#2>{\langle\,#1,\,#2\,\rangle}
\newcommand{\arr}{\begin{array}{rlll}}
\newcommand{\ea}{\end{array}}
\newcommand{\bea}{\begin{eqnarray}}
\newcommand{\eea}{\end{eqnarray}}
\newcommand{\bean}{\begin{eqnarray*}}
\newcommand{\eean}{\end{eqnarray*}}

\newcommand\Ricperp{\operatorname{Ric}^{\cD^\perp}}

\begin{document}
\title[Super-Poincar\`e algebras and  supergravities (I)]
{Super-Poincar\`e algebras, \\
space-times and  supergravities (I)}
%\arxurl{math-ph/1011.2722}
\author[A. Santi, A. Spiro]{$^{a}$A. Santi$^{*}$\footnote{*The first author was supported by project F1R-MTH-PUL-08HALO-HALOS08 of University of Luxembourg.} and $^{b}$A. Spiro}
\address{$^{a}$Facult\'e des Sciences, de la Technologie et de la Communication,\\
Universit\'e du Luxembourg,
L-1359 Grand-Duchy of Luxembourg\\
%\addressemail{}
\texttt{andrea.santi@uni.lu}}
\address{$^{b}$Scuola di Scienze e Tecnologie, \\ Universit\`a di Camerino,
Camerino, Italy\\
%\addressemail{}
\texttt{andrea.spiro@unicam.it}}
\begin{abstract}
A new formulation of theories of supergravity as theories satisfying a generalized Principle of General Covariance 
is given. It is  a generalization  of the superspace formulation of simple 4D-supergravity of  Wess and Zumino and it is designed to obtain geometric descriptions for  the supergravities that correspond to the super Poincar\`e algebras of  Alekseevsky and Cort\'es' classification. 
\end{abstract}
\maketitle
%\null \vspace*{-.25in}
\cutpage 
\setcounter{page}{2}
\noindent
\section{Introduction}
Up to now various   theories of supergravity, in diverse dimensions and based  on many super-extensions of Poincar\`e algebras, have been constructed. Although  super-extensions of Poincar\`e   algebras and algebras of Lorentzian symmetric spaces have been  already   classified under various   natural hypothesis   (see e.g. \cite{Na, Sr, AC, Sa}), to the best of our knowledge,  there does not exist a methodical  presentation of supergravity   theories that  parallels those lists of  super-extensions.  \par
\smallskip
We also recall that for gauge theories of classical Poincar\`e algebras, like  General Relativity,    the requirement of invariance under  localizations of  translations
 is  just  a re-formulation of the classical  Principle of General Covariance, i.e. the principle of  invariance under local changes of coordinates (or, equivalently, local diffeomorphisms) of the space-time (see e.g. \cite{Pi, St, ST, CDF}). By analogy,  it is natural to expect  that,  also for supergravity theories,  the  supersymmetries (analogues of  localizations of  translations) can be identified with Lie derivatives  along vector fields of an appropriate super-manifold and 
 that the requirement of  supersymmetric invariance  
  can be  stated as  a suitably  generalized Principle  of  General Covariance. \par
 \smallskip
On this regard, we would like to point out that  when a supergravity  can be  presented in a {\it manifestly covariant} way, i.e.  in terms of  tensorial equations,   the Principle of General Covariance is automatically satisfied and the off-shell invariance of the theory  is assured, with no need of explicit computations in coordinates or components.  \par
 \par
 \smallskip
The expectation that the invariance conditions of supergravity can be stated in terms of Lie derivatives  is supported  by the very  first superspace formulation of simple 4D-supergravity  (\cite{WZ}).  But an explicit and clear  formulation in such terms seems to us  still missing.   
So,  here and in \cite{SaS},  we  offer  a  presentation of  supergravities based on a generalized Principle of General Covariance and  involving 
a very small number of tensorial objects. \par
It  can be considered as a generalization of   the superspace formulation of Wess and Zumino:  As in \cite{WZ},  the physical fields   are presented as restrictions to  space-time  $M_o$ (not necessarily 4-dimensional)  of fields defined over   a superspace $M$, which has $M_o$ as a body,  and  the usual supersymmetries  are presented as   appropriate (infinitesimal)  local diffeomorphisms of  $M$. \par
\smallskip
Our  definitions  are designed so as to depend  in a canonical way on an initial  choice of a  super-extension $\gg$ of a Poincar\`e algebra.
We consider only the super-Poincar\`e algebras classified by Alekseevsky and Cort\'es  (\cite{AC})  corresponding to  
$N =1$  supergravities, but  the whole scheme can be  easily repeated 
for other super-algebras and  $N = p$ supergravities with $p \geq 2$.  
Notice also that our main  goal  was   to  reach   a simple and economical description 
of existing supergravity theories in terms of  objects  that can be studied with standard  techniques  of Differential Geometry. 
We {\it did  not}  address questions  on  the construction  of   Lagrangians, but we do expect interesting consequences   on this topic too.  \par
\smallskip
Here is a more  detailed description of our results.\par
\smallskip
In \S 2,  after recalling  some facts on  $\bZ_2$-graded and super-extensions of Poincar\`e algebras $\gg = \so(V) + V + S$   of  a pseudo-Riemannian space $V = \bR^{p,q}$,  
 we introduce the notion  of {\it space-time of type $\gg$}, which is   a (super) manifold $M$  with a  distinguished submanifold $M_o \subset M$ and  a non integrable distribution $\cD$, whose    Levi form $\cL$ is modeled on the Lie brackets of elements in  $S \subset \gg$.  Then we define as 
 {\it gravity field} any pair $(g, \n)$ formed by a   tensor field $g$ on $M$ of type $(0,2)$,   inducing a pseudo-Riemannian metric  on the  $g$-orthogonal distribution $\cD^\perp$ and by a covariant derivation    $\n$ preserving   $\cD$, $g$ and $\cL$. Properties of  these connections  are also given.\par
 \smallskip
In \S 3, we define as  {\it supergravity of type $\gg$} any pair formed by  a space-time  $(M, M_o, \cD)$ of type $\gg$ and  a gravity field $(g, \n)$. 
Any supergravity induces on  $M_o$  (which represents the space-time of Physics) the  following {\it physical fields}:  two covariant derivations,  called  {\it metric} and {\it spinor connections}, and three tensor fields, corresponding to the {\it graviton}, the {\it gravitino} and the {\it auxiliary field(s)}. Then we  state our generalized  {\it Principle of Infinitesimal General Covariance} and the notion of {\it manifestly covariance} for constraints and equations.   \par
\smallskip
In \S 4, we consider the class of {\it (strict) Levi-Civita supergravities of type $\gg$}, characterized  by the vanishing  of  certain parts of the  torsion $T$ of   $\n$.  The connections
satisfying these conditions  are generalizations of the  Levi-Civita connections of pseudo-Riemannian manifolds and {\it we prove for them   an   existence and uniqueness theorem}.  From this result it follows that
the physical fields of strict Levi-Civita supergravities
are completely determined by the graviton, the gravitino and the auxiliary field(s), as in supergravities formulated in the component approach.  Finally,   we determine  the transformation rules for the graviton, gravitino and the auxiliary field of a Levi-Civita supergravity. The expressions  nicely match the well-known rules of simple 4D-supergravity and other  supergravities.  \par
\smallskip
In  \S 5, we give  examples on how known theories of supergravity can be  presented as theories of Levi-Civita supergravities  of type $\gg$.\par
\smallskip
Our results  is  re-formulated  and formalized in  the language of supermanifolds  in  \cite{SaS}.  We chose to postpone such formalization in a second paper for the  following reasons.  
It is very common to deal  with supermanifolds in a naive way and consider them  just  as   smooth manifolds with points labeled by two kinds of coordinates,  the bosonic and the fermionic ones.  Following this habit,   we  give  here definitions and results  on   gauge theories of super and  non-super  extensions of Poincar\`e algebras,  with  proofs that can be  considered rigorous  only for what  concerns the latter and essentially correct for the former only  if   one consider  supermanifolds  ``as if''  they were  smooth manifolds.  In \cite{SaS},  we convert  everything  into rigorous statements on supermanifolds and on the gauge theories of super Poincar\`e algebras.  \par
\smallskip
Before concluding, we need to recall   that a presentation of supergravity, which is based  on a  Principle of General  Covariance,  appears also in the so-called  ``rheonomic approach''  of Regge, Ne'eman, Castellani, D'Adda, D'Auria, Fr\'e and van Nieuwenhuizen (see e.g. \cite{NR, NR1, DDFR, DFR, CDF}), where supergravities are described as theories of fields   on a soft-group manifold  $P$, a sort of principle bundle over the superspace $M$.   
We  also   remark  that  our approach
is crucially based on the notion of  the non-integrable distribution $\cD$ modeled on $\gg$:
To the best of our knowledge, similar non-integrable distributions have only  been considered  in the geometrical approach of  Ogievetsky, Sokatchev, Rosly\v\i, Schwarz et al. (see e.g. \cite{OS, OS1, OS2, OS3, Sc, RS, RS1, RKS,  Ma, Lo}) and in the superspace formulation of supergravity of  P. Deligne  (\cite{De}).  Analogies and differences  will be carefully discussed elsewhere. 
\par
\smallskip
\noindent{\it Notation.} Throughout  the paper,  we consider  Clifford algebras as defined   e.g. in \cite{LM}.  According to this,  the Clifford product  of vectors   of the standard basis of $\bR^{p,q}$ is  $e_i \cdot e_j = -  2 \eta_{ij}$ and  not  `` $+  2 \h_{ij}$'' as it is  often assumed   in  Physics literature.\par
\par
\smallskip
\font\smallit = cmti10
\noindent{\it Acknowledgements}. The authors are grateful to   D. V. Alekseevsky, V. Cort\'es,  C.\ Devchand  and M. Tonin for  helpful discussions and insightful comments.
\section{Space-times and gravity fields of type $\gg$}
\subsection{Extended Poincar\`e algebras and associated space-times}
Let  $V = \bR^{p,q}$  and    $\gp(V) = Lie(Iso(\bR^{p,q})) = \so(V) + V$   its {\it Poincar\`e algebra\/}. 
\begin{definition} \label{definitionextensions} A $\bZ_2$-graded Lie algebra (resp. a super-algebra) $\gg = \gg_0 + \gg_1$ is called {\it extended} (resp. {\it super}) {\it Poincar\`e algebra} if
%{\it  extension}  {\it of $\gp(V)$\/} if: 
\begin{itemize}
\item[a)] $\gg_0 = \gp(V) = \so(V) + V$; 
\item[b)] $\gg_1=S$ is   an irreducible   spinor module (i.e. an irreducible real representation of the Clifford algebra $\cC \ell(V)$ of $V$) and the adjoint action $\ad_{\so(V)}|_S:  S \longrightarrow S$ coincides with the standard action of $\so(V)$ on $S$ (i.e. $[A,s]  = A \cdot s$ for any $A \in \so(V)$, $s \in S$); 
\item[c)]  $[V, S] = 0$; 
\item[d)]  $[S, S] \subseteq V$.
\end{itemize}
%Any extension of $\gp(V)$ will be called  {\it extended Poincar\`e   algebra\/} (or {\it super-Poincar\`e algebra}).
  \par
 If $\gg$ is  an extended (resp. super) Poincar\`e  algebra,   any  connected homogeneous 
(super) space $M = G/H$, with   $Lie(G) = \gg$ and $Lie(H) = \so(V)$,  will be  called  {\it flat space-time of type $\gg$\/}.   The submanifold $M_o =  G_o/H \subset  M$, with  $G_o \subset G$  connected and   $Lie(G_o) =   \so(V) + V$,  is   called {\it body} of the space-time.\end{definition}
\medskip
As we have done in this definition, 
all   statements  and arguments of this paper have a ``super'' and a ``non-super''  version.  But,  hoping to be    clear and at the same time rigorous,  from now on {\it  we  give exact and precise definitions and statements    only for  the ``non-super''  case}.  Corresponding accurate definitions and  statements   for the  ``super''  case  
%also be given,  but the exact  meaning of the used terminology 
will be  given  in \cite{SaS}.  Nonetheless, it should not be hard to  understand their contents  on the base of  analogies.  \par
  \bigskip
We now want to introduce  a generalization  of  the notion of flat space-time,  which is fundamental in our presentation of supergravity theories. For this, we need to recall some notion, commonly used in studying   CR structures and  non-integrable distributions.  Let   $M$ be a manifold of dimension $m$ and  $\cD \subset TM$  a distribution of rank $p \leq m$ on $M$.  At any point $x \in M$, we may consider the map
\beq \label{Leviform} \cL_x : \Lambda^2 \cD_x \longrightarrow T_x M/\cD_x \ , \qquad \cL_x(v, w) = [X^{(v)}, X^{(w)}]_x \mod \cD_x\eeq
where $X^{(v)}$, $X^{(w)}$ are  vector fields in $\cD$  with $X^{(v)}_x = v$ and  $X^{(w)}_x = w$. A  simple check shows  that   $\cL_x(v, w)$  depends only on $v$ and $w$  and  that  \eqref{Leviform} is  a well-defined bilinear map. It  is  called  {\it Levi form  of $\cD$ at $x$\/}.  
\par
\medskip
 We say that  $\cD$ is  {\it of uniform type\/} if its Levi form $\cL_x$ is  independent on $x$ up to linear isomorphisms (i.e.  if for  any $x, y \in M$ there exists  an isomorphism $\imath: T_x M \overset\simeq\longrightarrow T_y M$  so that  
$\imath(\cD_x) = \cD_y$ and $\imath^*(\cL_y)  = \cL_x$).
\par
\medskip
\begin{example}
Any flat space-time $M = G/H$  is naturally endowed with   a $G$-invariant distribution, i.e. the unique invariant distribution  $\cD^\gg$ such that  $$\cD^{\gg}|_o= S\ ,\qquad o = eH$$
(we  use the standard identification $T_o G/H \simeq  V + S $). 
This distribution is of uniform type, transversal to the body $M_o = G_o/H$   and  with  Levi form at  $o$  
$$ \cL^\gg_o(s, s') =  [s, s']\ , \qquad s, s' \in S\ .$$
 \par
 \medskip
 If  $G/H$ is simply connected, $\cD^\gg$ is  described in    coordinates as follows. 
Let $ (e_i, e_\a)$  be a basis for $V + S$ with $e_i \in V$,  $e_\a \in S$.  The exponential map  $\exp: \gg \longrightarrow G$ induces a diffeomorphism 
\beq\label{flatexample} \exp: V + S \overset{\simeq}\longrightarrow G/H \eeq
and we may consider the  global system of coordinates $\xi: G/H \longrightarrow \bR^{\wh n}$, $\wh n = \dim V + \dim S$,  that associates to  any $x = \exp(x^i e_i + \theta^\alpha e_\alpha)$ 
the coordinates $ \xi(x) = (x^1, \dots, x^n, \theta^1, \dots, \theta^{\wh n - n})$.  \par
A vector $v = v^i e_i + v^\a e_\a \in V + S \simeq  T_o G/H$ is represented in the coordinate basis as $v = v^i \left.\frac{\partial}{\partial x^i}\right|_o + v^\a \left.\frac{\partial}{\partial \theta^\a}\right|_o$ and it is the tangent vector at $t = 0$ of the curve  $\g_t = \exp(t(v^i e_i + v^\a e_\a)) \in G/H$.  By BCH-formula, an element $g =  \exp(x^j e_j + \theta^\b e_\b) \in \exp(V + S)\subset G$ maps 
$\g_t$ into the curve
$$g \cdot \g_t = \exp(x^j e_j + \theta^\b e_\b + t(v^i e_i + v^\a e_\a) +\frac{1}{2} t  v^\alpha \theta^\b \cL_{\b \a}^k e_k)\ ,$$
 where  $\cL_{\a\b}^i$ are the components of the Levi form $\cL^\gg_o$ in the basis $(e_i, e_\a)$. From this it follows  that  
$$g_*(v) = v^i \left.\frac{\partial}{\partial x^i}\right|_{(x^j, \theta^\b)} +  v^\a\left( \left.\frac{\partial}{\partial \theta^\a}\right|_{(x^j, \theta^b)} + \frac{1}{2}  \theta^\b  \cL_{ \b\a}^i \left.\frac{\partial}{\partial x^i}\right|_{(x^j, \theta^\b)} \right)$$
and hence  that any  linear combination of the vector fields 
\beq \label{2.8}E_i \=ÿ\frac{\partial}{\partial x^i}\ ,\qquad E_\a \= \frac{\partial}{\partial \theta^\a} +\frac{1}{2}  \theta^\b \cL_{\b\a}^i\frac{\partial}{\partial x^i}\eeq
is $\exp(V + S)$-invariant (see e.g.  \cite{We}, Ch. 14). Finally,  the $G$-invariant distribution $\cD^{\gg}$ of $G/H$ is generated by the fields $E_\a$, i.e. 
$\cD^\gg_{x} = \Span_\bR\left\{ E_\a|_{x}\right\}$.
%$$\cD^\gg_{x} = \Span_\bR\left\{ E_\a|_{x}\right\} \qquad \text{for any}\ \ x = \xi^{-1}(x^j, \theta^\b) \in G/H\ .$$
\end{example}
\par
\medskip
The properties of $\cD^\gg$ of previous example motivate the  following notion.
\par
\begin{definition} 
\label{spacetime}
A  {\it space-time of type $\gg$\/} is any triple $(M, M_o, \cD)$ given by: 
\begin{itemize}
\item[--]  a connected manifold $M$ of dimension $\wh n =
\dim V + \dim S$; 
\item[--]  
a connected submanifold $M_o \subset M$  of dimension $
n = 
\dim V$; 
\item[--] a distribution   $\cD \subset TM$ of rank $
n^S =  
\dim S$ and transversal to $M_o$ (i.e. with $T_x M_o \cap \cD_x = \{0\}$ at any $x \in M_o$) satisfying the following  ``uniformity assumption'': 
\begin{itemize}
\item[]
 {\it for any $x\in M$  there exists a neighborhood $\cU \subset M$ of $x$ and a smooth family of vector space  isomorphisms  
$\imath^{(y)}:  V+S\longrightarrow T_y M$, $y \in \cU$, so that 
$$\imath^{(y)}(S)=\cD_x \qquad \text{and}\qquad 
\imath^{(y)}{}^*(\cL_y) =  \cL^\gg_o\ ;$$ if   $S=S^+ + S^-$  is sum of irreducible $\so(V)$-moduli, we also assume 
$\cD = \cD^+ + \cD^-$ for distributions $\cD^{\pm}$ and  $\imath^{(y)}(S^{\pm})=\cD^{\pm}_y$.}
\end{itemize}
\end{itemize} 
The submanifold $M_o$ is called {\it body} of the space-time.	
\end{definition}
%In case $S=S^+ + S^-$ is a sum of irreducible $\so(V)$-moduli, we call space-time of type $\gg$ any triple $(M, M_o, \cD = \cD^+ + \cD^-)$, where $\cD^{\pm}$ are two distributions so that $\imath_x(S^{\pm})=\cD^{\pm}_x$ for an isomorphism $\imath_x$ as above.\par
\medskip
Notice that, if    $(M, M_o, \cD)$ is a (non-flat) space-time  and  $(E_\a)$  is a set of local generators for $\cD$ around a point $x_o \in M_o$, then
it is always possible to determine  a  system of coordinates 
$(x^i, \theta^\a)$ on a neighborhood $\cU$ of $x_o$ so that 
$$
\label{coordinateadattate} M_o \cap \cU = \{\ \theta^\a = 0\ \}\ ,\qquad E_\a|_{M_o} = \left.\frac{\partial}{\partial \theta^\a}\right|_{M_o}\ ,$$
as it occurs on the  flat space-time considered above. On the other hand, it goes without saying that the expressions for the
 $E_\a$'s outside the body $M_o$ are in general quite different from the (\ref{2.8}).
\subsection{Admissible extended (or super) Poincar\`e algebras and associated gravity fields}
\label{subsection3.2}
\subsubsection{Admissible extended and admissible super  Poincar\`e algebras}
In \cite{AC} it was observed that, given an irreducible spinor  module $S$,   any   extension $\gg = \so(V) + V + S$ of $\gp(V)$   is completely determined by the tensor 
$$L \in  \Lambda^2 S^* \otimes V\qquad \text{(resp.} \ \ \vee^2 S^* \otimes V\ )$$
 that  defines  the Lie brackets $[\cdot ,\cdot ] $ between   elements  in $S$. 
This tensor  is $\so(V)$-invariant and any $\so(V)$-invariant tensor  of this kind  corresponds to a unique structure of extended (or super) Poincar\`e  algebra on $ \so(V) + V + S$. \par
\medskip
A tensor   $L \in \Lambda^2 S^* \otimes V$ or $\vee^2 S^* \otimes V$  is called {\it admissible} if the  associated tensor  
$$L^*  \in S^* \otimes S^* \otimes V^*\ ,\qquad L^*(s, s', v) \=  <L(s,s'), v>$$ 
is of the form
\beq\label{admissible}L^*(s,s', v) = \beta(v\cdot s, s')\ ,  \eeq
for some  non-degenerate $\so(V)$-invariant bilinear form  $\b$  on $S$
%where ``$\cdot$'' is the Clifford action of $V$  on   $S$, and  
such that: 
\begin{itemize}
\item[1)] it is  either symmetric or skew-symmetric; 
\item[2)] the Clifford multiplications $ v \cdot ( \cdot): S \longrightarrow S$, $v \in V$,  are either  all $\b$-symmetric or all $\b$-skew symmetric; 
\item[3)] if $S$ is sum  of irreducible $\so(V)$-moduli $S = S^+ + S^-$, then $S^\pm$ are either mutually $\b$-orthogonal or both $\b$-isotropic. 
\end{itemize}
Any admissible tensor is $\so(V)$-invariant, it corresponds to an extended (or super) Poincar\`e  algebra and  {\it the spaces $(\Lambda^2 S^* \otimes V)^{\so(V)}$ and $(\vee^2 S^* \otimes V)^{\so(V)}$ have  bases of  admissible elements\/} (\cite{AC}).  \par
\medskip
\begin{definition} \label{definnerproduct}
An extended (or super) Poincar\`e  algebra $\gg = \so(V) + V + S$ is called {\it admissible} if it is determined by an admissible tensor $L$. 
In this case,  if $\b$ is the bilinear form \eqref{admissible}, we call
 {\it extended  inner product of $V + S$} the non-degenerate  bilinear form $(\cdot, \cdot)$, defined by 
\beq \label{innerproduct} (\cdot, \cdot)|_{V \times S} = 0\ ,\quad (\cdot, \cdot)|_{V \times V} =  <\cdot ,\cdot >\ ,\quad (\cdot, \cdot)|_{S \times S} =\b\ .\eeq
\end{definition}
\par
\medskip
From now on,   {\it any extended (super) Poincar\`e algebra  will be  assumed to be admissible\/} and $(\cdot, \cdot)$  will always indicate the bilinear form \eqref{innerproduct}. (\footnote{
Actually, for many of  our  results, it is sufficient to consider a non-degenerate $\so(V)$-invariant bilinear form \eqref{innerproduct}, with $\beta$ not necessarily equal to the one in \eqref{admissible}.
%This assumption is mainly for simplicity. For most of the following results, it is sufficient to assume the existence of a $\so(V)$-invariant non-degenerate bilinear form $\beta$ on $S$
})
\par
\begin{example} \label{4Dexample}
Let $V=\bR^{3,1}$ and denote by  $(e_0,\dots,e_{3})$ its standard basis  with  $< e_i, e_j>= \varepsilon_i\d_{ij}$ with  $\varepsilon_0=-1$, $\varepsilon_1=\varepsilon_2=\varepsilon_3=+1$. 
Let  also $S=\bC^4$ and denote by 
$\rho:\cC\ell_{3,1}\longrightarrow\End(S)$ the Dirac representation of $\cC\ell_{3,1}$, determined by the $\G$-matrices
$$\G_0 = \rho(e_0) = \left( \begin{matrix} 0 & I \\
I & 0 \end{matrix}\right) \ ,\qquad \G_i = \rho(e_i) = \left( \begin{matrix} 0  & \s_i \\
- \s_i & 0  \end{matrix}\right) \ ,\qquad i = 1,2,3\ $$
where the $\s_i$ are the usual Pauli matrices $\s_1 =  \left( \smallmatrix 0 & 1 \\ 1 & 0\endsmallmatrix\right)$, $ \s_2 = \left(
\smallmatrix 0 & -i \\ i & 0\endsmallmatrix \right)$, $\s_3 = \left(
\smallmatrix 1 & 0 \\ 0  & -1\endsmallmatrix \right)$.
Let also 
 \beq  \label{gammafive} \G_5\= i \G_0 \G_1 \G_2 \G_3 = \left( \begin{matrix} -I  & 0 \\
0 & I \end{matrix}\right)\eeq and  
$S = S^+ + S^- = \bC^2 + \bC^2$ be the corresponding  decomposition of $S$ in $\G_5$-eigenspaces, i.e.  into irreducible $\so(V)$-moduli of Weyl spinors, on which $\so(V)$ 
acts by conjugate  representations.
Finally, let  $\varepsilon$ be  the standard volume form of $ \bC^2 = S^+ = S^-$ and  $\o \in  \Lambda^2 S \simeq\L^2 \bC^4$
the 2-form 
\be \label{volumeform}  \o(s ,s') = \varepsilon(s^+, s'{}^+) - \varepsilon(s^{-}, s^{'-}) = s^T  C s'\ ,
\ee
where we considered the decompositions $s = s^+ + s^-$,  $s' = s^{'+} + s^{'-}$  into $S^{\pm}$- components and $C = -  i \G_0 \G_2$ is the 
charge conjugation matrix.\par
The admissible  bilinear forms
$$\beta_1(s, s')= \Re \o(s, s) = - \Re(is^T \G_0\G_2 s')\ ,$$
$$\beta_2(s, s')=\Im \o(s, s') = - \Im(is^T \G_0 \G_{2} s')\ , $$
$$\label{basissuper} \beta_3(s, s')=\Re(\overline{s}^T \G_0 s')\ , \qquad \beta_4(s, s')=\Re(\overline{s}^T\G_5\G_0  s')\ .$$
give a  basis for the space of tensors associated with   {\it super} extensions of $\gp(\bR^{3,1})$, while 
%\footnote{In Physics,  given a spinor $s$, the cospinor  $\wt s \overset{def}= \overline{s}^T\G_0$ is called {\it Dirac conjugate of $s$}.}.  
%They are  a basis for the $\so(V)$-invariant tensors associated with  super-extensions of  $\gp(\bR^{3,1})$. 
the admissible  bilinear forms
$$\wt\beta_1(s, s')=\Im(s^T \G_1\G_3 s')\ ,\qquad \wt \beta_2(s, s')=\Re(s^T \G_1 \G_{3} s')\ , $$
$$\wt\beta_3(s, s')=\Im(\overline{s}^T\G_0 s')\ , \qquad\wt\beta_4(s, s')=\Im(\overline{s}^T\G_5\G_0  s')\ ,$$
give a basis for the space of tensors associated with   {\it non super} extensions. 
%the space of $\so(V)$-invariant $\bC$-bilinear forms is given by
%$$
%\beta(s,s')=is^T\G_0\G_2 s'\,\,\,\,\,\,\,\,,\,\,\,\,\,\,\,\,\,\beta(s,s')=s^T\G_1\G_3s'
%$$
%so that the 
\end{example}
\begin{example} \label{11Dexample}
Let $V = \bR^{10, 1}$ and again  denote by $(e_0, \dots, e_{10})$ its standard basis with  $< e_i, e_j> =  \varepsilon_i \d_{ij}$ with  $\varepsilon_0=-1$, $\varepsilon_i=+ 1$ for $1\leq i \leq 10$. Let $S =  \bC^{32}$ and 
$\rho:\cC\ell_{10,1}\longrightarrow \End(S)$ the Dirac representation of $C\ell_{10,1}$, determined by purely imaginary $\Gamma$-matrices  $\G_i=\rho(e_i)$ (see e.g. \cite{MS}). 
The admissible  bilinear forms 
$$ \beta_1(s, s')=\Re(is^T \G_0 s')\ ,\ \  \beta_2(s, s')=\Im(is^T \G_0 s')\ ,\ \  
\beta_3(s, s')=\Re(\overline{s}^T \G_0 s')\ ,
$$
give a  basis for the space of tensors associated with  {\it super} extensions of $\gp(\bR^{10,1})$, 
while the admissible bilinear form
$$\beta(s, s')=\Im(\overline{s}^T\G_0 s')$$
is a basis for the space of  tensors associated with {\it non super} extensions. 
%(see again \cite{AC}, Table 9).
\par
%A basis of the space of $\so(V)$-invariant $\bC$-bilinear forms is given by
%$$
%\beta(s,s')=is^T\G_0 s'
%$$
\end{example}
%\begin{example} \label{11Dexample}
%Let $V = \bR^{1,10}$ and denote by $(e_0, \dots, e_{10})$ its standard basis with  $< e_i, e_j> =  \varepsilon_i \d_{ij}$ with  $\varepsilon_0=+1$, $\varepsilon_i=- 1$ for $1\leq i \leq 10$. Let $\mathbb{S} =  \bC^{32}$ and 
%$\rho:\cC\ell_{1,10}\longrightarrow \End_{\bC}(\mathbb{S})$ the Dirac representation of $C\ell_{1,10}$,
%\simeq \bR(32)\oplus\bR(32)$
%determined by purely real $\Gamma$-matrices  $\G_i=\rho(e_i)$ (see e.g. \cite{MS}). The 
%irreducible real representation $S$ of $\cC \ell_{1,10}$ is $\bR^{32}\subset\bC^{32}$ and
%the admissible bilinear form
%\beq \beta(s, s')=\Re(s^T \G_0 s')\ .
%\eeq
%is, up to scalars, the only $\so(\bR^{1,10})$-invariant bilinear form of $S$. It gives a basis for the space of tensors associated with the  {\it super} extensions of $\gp(\bR^{1,10})$. 
%\par
%A basis of the space of $\so(V)$-invariant $\bC$-bilinear forms is given by
%$$
%\beta(s,s')=is^T\G_0 s'
%$$
%\end{example}
\subsubsection{Gravity fields of type $\gg$}
%We now introduce the notion of gravity field. In case "$S = 0$", it is just a   pair $(g,\wt  \n)$ formed by a pseudo-Riemannian metric $g$ and a metric connection $\wt \n$ on $M_o$.  It is  therefore a generalization of the  objects (i.e. metric and Levi-Civita connection) representing gravitational forces in General Relativity.\par
In the following definition, we denote by $\gg$  an admissible extended Poincar\`e  algebra with extended inner product $(\cdot, \cdot)$ and 
by  $(M, M_o, \cD)$  a space-time of type $\gg$ with Levi form $\cL$. 
\begin{definition}\label{gravityfields2}
  A  {\it  gravity field on $(M, M_o, \cD)$\/} is a  pair $(g, \n)$  formed  by a tensor field $g$ of type $(0,2)$   and a connection $\n$ on $M$  so that: 
\begin{itemize}
\item[i)] the tensor $g$ is so that, for any $x \in M$, there exists  a neighborhood $\cU \subset M$ of $x$ and a smooth family of vector space  isomorphisms  
$\imath^{(y)}:  V+S\longrightarrow T_y M$, $y \in \cU$, so that: 
\begin{itemize}
\item[a)]  $\imath^{(y)}(S) = \cD_y$, $\imath^{(y)}(V) = \cD^\perp_y$ and,  if  $S = S^+ + S^-$, $\imath^{(y)}(S^\pm) = \cD^\pm_y$;
\item[b)]  $ \imath^{(y)}_*(\cdot, \cdot) = g_y $;
\item[c)] $\imath^{(y)}_*(\cL^\gg_o) = \cL^{g}_y$, where $ \cL^{g}_y \in \Hom(\cD_y\times \cD_y, \cD^\perp_y)$ is 
$$\cL^g_y\= (\pi|_{\cD^\perp})^{-1} \circ \cL_y$$ 
and   $\pi|_{\cD^\perp}: \cD^{\perp}  \longrightarrow T M /\cD$ is the natural isomorphism between the $g$-orthogonal distribution $\cD^\perp$ to $\cD$ and the bundle $T M /\cD$; 
\end{itemize}
\item[ii)]  the distribution $\cD$ is $\n$-invariant  and, if $S = S^+ + S^-$, the distributions $\cD^\pm$ are $\n$-invariant; 
\item[iii)]   $\n g = 0$ and  $\n \cL^g = 0$. 
\end{itemize}
%In case $S=S^+ +S^-$ is sum of two irreducible semi-spin modules, a gravity field is assumed to satisfy also 
%\begin{itemize}
%\item[iv)] the distributions $\cD^{\pm} \subset \cD$,  given by the spaces $\cD^{\pm}_y = \imath^{(y)}(S^\pm)$,  where $\imath^{(y)}$ an isomorphisms described in (i), are 
%$\wt \n$-stable. 
%\end{itemize}
In this case, we say that $g$ is  the {\it extended metric} and $\n$ the  {\it extended metric connection}. 
\end{definition}
\smallskip
The name ``extended metric''   stems from  the notion of ``extended inner product'' (see Definition \ref{definnerproduct}) and  one should  keep in mind  that $g$ is not always a symmetric tensor field. Notice also that, from  (ii) and (iii),  any extended metric connection $\n$ preserves also the complementary distribution  $\cD^\perp$. \par
\bigskip
Let $(g, \n)$ be a gravity field on a space time $(M, M_o, \cD)$ of type $\gg$. We call {\it bundle of orthonormal  frames} of $g$
the collection $\operatorname{O}_g(M, \cD)$ of all vector spaces isomorphism $\imath: V + S \longrightarrow T_x M$ satisfying (a) - (c) of previous definition. 
Using (i), one can check that $\operatorname{O}_g(M, \cD)$ is indeed a principal bundle over $M$ with a structure group $G$, whose  identity component $G^0$
is the subgroup of $\GL(V + S)$
$$G^0 =\left \{\  \left(\begin{matrix}
k & 0  \\
0 & k\circ h
\end{matrix}\right),\ k \in \Spin^0(V)\ , h\in H^0\ \right\} = \Spin^0(V)\cdot H^0\ ,$$
 where $H^0$ is the identity component  of $H = \operatorname{O}(S,\beta)\cap C_{\ggl(S)}(\cC \ell(V))$ or of the  subgroup  of $H$, which   preserves $S^+$  and $S^-$,  when $S = S^+ + S^-$.\par
By definitions, the extended metric connection $\n$ preserves $\operatorname{O}_g(M, \cD)$ and it can be considered as 
the covariant derivation  on $M$ determined by a connection form $\o$ on 
 $\operatorname{O}_g(M, \cD)$. \par
\medskip
Consider now  the connections  $\n^o$ and $\n^{o'}$ induced by $\n$ on the vector bundles
$$
\pi: \cD^\perp \longrightarrow M\ ,\qquad 
\pi': \cD \longrightarrow M\ .
$$
They may be considered as the covariant derivations determined by 
 connection forms $\o^o$, $\o^{o'}$ on the bundles $\operatorname{O}_g(\cD^\perp)$ and $\operatorname{O}_g(\cD)$ of the   $g$-orthonormal frames of the spaces $\cD^\perp_x \subset T_x M$ and
$\cD_x \subset T_x M$, respectively. On the other hand, using  a (local) field of  frames $\imath^{(y)}: V + S \longrightarrow T_y M$, one can identify any space $\cD_y$ with the spinor module  $S$ and identify (at least locally)  the bundle
$\pi': \cD\longrightarrow M$ with the  spinor bundle associated with  $\operatorname{O}_g(\cD^\perp)$, i.e.
\beq \label{identification} \cD \simeq \Spin_g(\cD^\perp) \times_{\Spin^0(V)} S\ .\eeq
\par
\smallskip
For a fixed (local) identification \eqref{identification}, we may consider  on $\cD$  the covariant derivation induced by the covariant derivation $\n^o$ of $\cD^\perp$ (be aware that the induced derivation  depends on  the identification  \eqref{identification}). If we denote also this covariant derivation  by $\n^o$,  we have that 
\beq \label{tensorC} \n^{o'}_X s = \n^o_X s + C_X (s)\ ,\qquad X \in \gX(M)\ ,\  s \in \G(\cD)\ ,\eeq
for some  field  $C$   in $T_x^*M \otimes \cD^*_x \otimes \cD_x \simeq (V+S)^* \otimes S^* \otimes S$ at any $x \in M$. \par
\smallskip
In particular, $\n$ can be locally written as a sum of the form  
$\n = \n^o + C$, 
where  $C$ is defined  in \eqref{tensorC} and $\n^o$ is   sum of the connection on $\cD^\perp$ and the induced connection on $\cD$. Note that  $\n^o$  satisfies (ii), (iii) of Definition \ref{gravityfields2}. \par
\smallskip
As we pointed out above, such decomposition (and the field $C$) depends in principle on the chosen identification \eqref{identification}. But the next proposition shows that in many cases $C$ is   trivial, no matter what is  the used identification. 
\begin{prop}  \label{cruxunicity} Let $(g, \n)$ be a gravity field on $(M, M_o, \cD)$ and $\n = \n^o + C$ a decomposition determined 
by an identification \eqref{identification}. For any $x \in M$, the tensor $C_x$  belongs to   $(V+ S)^* \otimes \gh$, where  $\gh=Lie(H)$ is  contained in one  of the  subspaces of  $C_{\ggl(S)}(\cC \ell(V))$ described  in  Table 1:  
\vskip 0.5 cm
\centerline{
\tiny
\vbox{\offinterlineskip
\halign {\strut
\vrule \vrule \vrule\hfil\  $#$\  \hfil
&\vrule  \vrule\vrule\hfil\  $#$ \ \hfil
&\vrule \vrule\hfil\  $#$\ \hfil
& \vrule \vrule\hfil\  $#$\ \hfil
& \vrule \vrule \hfil\  $#$\ \hfil
& \vrule \vrule\hfil\  $#$\ \hfil
&\vrule \vrule \hfil\  $#$\ \hfil
&\vrule \vrule \hfil\  $#$\ \hfil
&\vrule \vrule \hfil\  $#$\ \hfil
\vrule \vrule
 \cr
\noalign{\hrule \hrule}
\underset{\phantom{A}}{\overset{\phantom{A}}{p-q \mod 8}} &
0 &
1 &
2 &
3 &
4 & 
5 &
6 &
7 \cr
\noalign{\hrule\hrule}
\underset{\phantom{A}}{\overset{\phantom{A}}{C_{\ggl(S)}(\cC \ell(V))} }
& \bR & \bC  & \bH & \bH &\bH & \bC & \bR & \bR  \cr
\noalign{\hrule\hrule}
\underset{\phantom{A}}{\overset{\phantom{A}}{
\smallmatrix irr.\ \so(V)\text{moduli}\\ \text{in}\ S\endsmallmatrix
} }
& S^+ \not\simeq S^-  & S^+ \simeq S^-  & S^+ \simeq S^-  & S & S^+ \not\simeq S^- & S & S & S  \cr
\noalign{\hrule\hrule}
\underset{\phantom{A}}{\overset{\phantom{A}}{\gh\ \text{is contained in}} } & 0 & 0  & \Span_\bR\{ i \} 
& \Span_\bR\{i, j, k \}  &\Span_\bR\{i, j, k\}  & \Span_\bR\{i\} & 0 & 0  \cr\noalign{\hrule\hrule}
}}
}
\vskip - 0.1 cm
\centerline{\tiny\bf Table 1}
In particular,   $\n = \n^o$ when $ p - q = 0,1,6, 7 \mod 8$. 
 \end{prop}
\begin{pf} By the properties of  $\n$ and $\n^o$,  for any vector fields $X,  v, s, s' \in \gX(M)$, with $v_x \in \cD^\perp_x (\simeq V)$ and $s_x, s'_x \in \cD_x (\simeq S)$ at all points,   we have that
$$g(v, \cL^g(C_X(s), s') + g(v, \cL^g(s, C_X(s')) = 0\ , \ \  g(C_X(s), s')  + g(s, C_X(s')) = 0\ .$$
Hence, using the identifications $T_xM \simeq V + S$ and the admissibility of $\gg$,  we get  that for any $X \in V + S$, $v \in V$ and $s, s' \in S$
$$\b(	v \cdot C_X(s), s' ) +  \b(v \cdot s, C_X( s')) = 0 \ ,\quad \b(C_X(v \cdot s), s') + \b(v \cdot s, C_X( s')) = 0\ .$$
By non degeneracy of $\b$, these conditions are equivalent to
$$ v \cdot C_X(\cdot) = C_X(v \cdot (\cdot))\ ,\, \quad  \b(C_X( \cdot),\cdot ) + \b(\cdot, C_X( \cdot)) =0\ ,$$
i.e.  $C_X\in \go(S, \b)\cap C_{\ggl(S)}(\cC \ell(V))$ (in addition, 
if  $S = S^+ + S^-$, the conditions  on $\n$ and $\n^o$ imply that $C_X$  preserves $S^+$ and $S^-$). \par
For any given signature $s = p-q$, the  centralizer $C_{\ggl(S)}(\cC \ell(V)) $ is immediately determined recalling that $\cC \ell(V) \simeq \bK(N)$ or $\bK(N) \oplus \bK(N)$ for some suitable $N$, with $\bK = \bR$, $\bC$ or $\bH$. In all cases, one can determine  the $\so(V)$-moduli in $S$ and the  elements in $C_{\ggl(S)}(\cC \ell(V)) $ that preserve these moduli   (see \cite{AC}, Prop. 1.5 and \cite{Fi}, Tables 1 and 2). Excluding  the elements which are real multiples of the identity (which cannot be in $\go(S, \b)$), one gets  the spaces listed in the last row of Table 1.
\end{pf}
\medskip
 \begin{rem} \label{Ctriviality} It should be stressed  that  Table 1 gives just  an upper bound for   $\dim \gh$. When  $\b$ is explicitly given, one gets a finer result   by direct computations. 
 %For instance, in the 4-dimensional supergravity of Wess and Zumino the form $\b$ is so that $\gh = 0$ and $\wt \n  = \n^o$ (see \S \ref{representing}).
 \end{rem}
% \smallskip
%For simplicity, in  the following we will always assume that  {\it a  gravity fields  admits a (local)   identification 
%$\cD \simeq \Spin_g(\cD^\perp) \times_{\Spin^0(V)} S$, for  which $\wt \n = \n^o$}. \par

\section{Theories  of  supergravity}
%\setcounter{equation}{0}
%\medskip
\subsection{Gravities and supergravities}
\label{gravitiesandsupergravities}
\begin{definition}\label{supergravityfields}
Let  $M_o$ be a manifold of dimension $n = \dim V$. We call {\it (super) gravity of type $\gg$ on $M_o$\/}  any pair $\cG= ((M, M_o, \cD), (g, \n))$ formed by 
\begin{itemize}
\item[a)] a  space time $(M, M_o, \cD)$ of type $\gg$ with body $M_o$; 
\item[b)] a gravity field $(g, \n)$  on $(M, M_o, \cD)$.
\end{itemize}
\end{definition}
Given a (super) gravity $\cG = ((M, M_o, \cD), (g, \n))$,  we call 
{\it spinor bundle of  $\cG$\/} the pullback bundle 
$$\pi : \cS = \cD|_{M_o} \longrightarrow M_o\ .$$
We also call  {\it physical fields of $\cG$\/}  the  following objects: 
 \begin{itemize}
  \item[--] the  tensor field in $T^*M_o \otimes_{M_o} \cS$, called  {\it gravitino\/}, defined by
\beq \label{31} \vartheta(X) \= \pi^{\cD}(X)\ ,\eeq
where, for any $x \in M$, we denote by    $\pi^{\cD}_x : T_x M \longrightarrow \cD_x$ the $g$-orthogonal projection onto $\cD_x$; 
 \item[--]  the tensor field in $\vee^2 T^* M_o$,  called  {\it graviton\/}, defined by 
 \beq\label{32}ÿ\phantom{aaaaaa} \wh g(X, Y)  \=  g(X, Y) -  g(\vartheta(X), \vartheta(Y)) = g(\pi^{\cD^\perp}(X),\pi^{\cD^\perp} (Y) )\ ,\eeq 
where, for any $x \in M$, we denote by  $\pi^{\cD^\perp}_x : T_x M \longrightarrow \cD^\perp_x$ the $g$-orthogonal projection onto $\cD^\perp_x$;
\item[--] the tensor field in  $ T^*M_o \otimes_{M_o} \cS^* \otimes_{M_o} \cS$, called  {\it A-field}, 
defined by 
\beq\label{33}ÿ \bA_{X s} \=   -  \pi^{\cD} ( T _{X s})\ , \eeq
where we denoted by $T$  the torsion of the connection $\n$; 
\item[--] the connection $D: \gX(M_o) \times \gX(M_o) \longrightarrow \gX(M_o)$, called  {\it metric connection\/}, defined by (\footnote{
Notice that  $D_X Y$ is  equal to    the projection  of $\n_XY$ onto $TM_o$  w.r.t. to the 
decomposition $TM|_{M_o}ÿ= TM_o + \cD|_{M_o}$. In other words,  $D$ is the connection on the submanifold $M_o \subset M$, induced by $\n$,  by  identifying 
the normal bundle 
$TM|_{M_o}/TM_o$ with  $ \cD|_{M_o}$.}) 
\beq \label{metconn}  D_X Y \= (\left.\pi^{\cD^{\perp}}\right|_{T M_o})^{-1}\left(\n_{X}\  \pi^{\cD^{\perp}}(Y)\right)\ ; \eeq
%determined by $g_x$
%and $  (\pi^{\cD^{\perp}}\circ \imath_*)^{-1}$ is inverse of  $\pi^{\cD^{\perp}}\circ \imath_*: T_x N \longrightarrow \cD^\perp_{\imath(x)}$; 
\item[--]   the  connection $\bD: \gX(M_o) \times \G(\cS) \longrightarrow \G(\cS)$ on the space $\G(\cS)$ of the sections of   $\pi: \cS \longrightarrow M_o$,  called {\it spinor connection\/},  defined by 
\beq 
\label{spinorconn}
\bD_X s \=  \n_{X} s  -  \pi^{\cD} ( T_{X s}) = \n_X s + \bA_X s\ .\eeq 
\end{itemize}
Finally, we call {\it non-physical  fields of $\cG$\/} the  tensor fields in $\cS^* \otimes_{M_o} \cS^* \otimes_{M_o} \cS$ and 
$T^* M_o \otimes_{M_o} \cS^* \otimes_{M_o} \cS^* \otimes_{M_o} \cS$, called {\it B-field} and {\it C-field}, 
defined by 
 \beq \label{B-field} \bB_{s s'} \= - \pi^\cD \left(T_{s s'}\right)\ ,\qquad  \bC_{X s s'} \= - \pi^\cD \left((\n_{s'}T)_{X s}\right)\ .\eeq
 \par
 \medskip
\begin{rem} Our presentation of supergravity theories is essentially based on this  definition and the contents of next subsection. 
In \S 5,  we will indicate how {\it various $N = 1$ supergravities  can be presented as theories on physical fields of 
supergravities of type $\gg$}.  
%The objective we have in mind  is 
%to put up a common framework for  the gauge theories of 
%super Poincar\`e algebras, in which various techniques of modern Differential Geometry can be used.
\end{rem}
\medskip
As it is suggested by our choice of names, the above defined  ``graviton'' and ``gravitino''  are precisely the objects, which we want 
to use   to formalize the common notions of graviton and gravitino in standard  supergravity theories.  \par
In fact,  from Definition \ref{gravityfields2}, one can check
that {\it the graviton $\wh g$ is a pseudo-Riemannian metric  of signature $(p,q)$ and
that   $D\wh g = 0$}. 
On the other hand,  given  a fixed  orthonormal  basis 
$(e_i, e_\a)$ for $(V + S, (\cdot, \cdot))$ and  a  corresponding local  frame field 
$$\left(\ E_i(x) = \imath^{(x)}(e_i)\ ,\ \  E_\a(x) = \imath^{(x)}(e_\a)\ \right)\ ,\qquad \imath^{(x)}\in \operatorname{O}_g(M, \cD)|_{M_o}\ ,$$
{\it  the field $\vartheta$  is of the form $\vartheta = \psi^\a_i E_\a \otimes E^i|_{TM_o}$, as the usual 
 gravitino}  (see  \cite{WB}). 
%\medskip
%We will see in \S 6 that such interpretations of the notions of ``graviton'' and ``gravitino'' 
% are consistent w.r.t. to the  transformation rules under localized super symmetries, adopted 
% in the most classical  supergravity theories. \par
\subsection{The Principle of General Covariance and manifestly covariant equations}%\hfill\par
\label{generalcovariance}
As we mentioned in the Introduction,   we want to  present the  transformation rules of a supergravity theory  as  actions of infinitesimal diffeomorphisms (= Lie derivatives) 
%of a supergravity $(M, M_o, \cD,g,\wt \n)$ 
and  generalize the Principle of General Covariance. 
% (\footnote{\ In \cite{ST}, the Principle of General Covariance is called ``Equivalence Principle''.}). 
\par
\medskip
We first remark that  for any (super) gravity $\cG= ((M, M_o, \cD), (g, \n))$    of type $\gg$,  any (local) diffeomorphism  $ \varphi:   M \longrightarrow M$,   sufficiently close to  $\Id_M$, determines a new a pair
\beq \label{actionofdiffeomorphisms}ÿ\cG' =  \varphi_*(\cG) \= ((M, M_o, \varphi_*(\cD)),((\varphi^{-1})^{*} g, ( \varphi^{-1})^{*}\n))\eeq 
which still is a (super) gravity  of type $\gg$. %\begin{pf} If $\wh \varphi$ is sufficiently close to $\Id_M$, the distribution $\cD'$ is still  transversal to $M_o$ and,  from the invariance of the Levi form under diffeomorphisms,  $\cD'$ is of uniform type and $(M, M_o, \cD')$ is an extended space time.  One can also check that $(\wh \varphi_* g, \wh \varphi_* \wt\n)$ is a gravity on $(M, M_o, \cD')$ and that $\wh \varphi_*(\cG)$ is a (super) gravity of the same type of $\cG$. 
%\end{pf}
%\begin{definition} 
%Consider two (super) gravities  
%$\cG = ((M, M_o, \cD), (g, \wt \n))$ and $ \cG' = ((M, M_o, \cD'), (g', \wt \n'))$ of  type $\gg$. We say that they are: 
%\begin{itemize}
%\item[--] {\it fully equivalent\/} if  the spinor bundle $\cS = \left.\cD\right|_{M_o}$ is equal to the spinor bundle $\cS' = \left.\cD\right|_{M_o}$ and  if  they determine the same physical fields   over $M_o$; 
%\item[--]   {\it weakly diffeomorphic\/} if there exists a  fiber bundle equivalence $\varphi:  \cS \longrightarrow \cS'$  mapping  
% the physical fields  of $\cG$ 
%into the corresponding physical  fields of $\cG'$; 
%\item[--]   {\it  (strongly) diffeomorphic\/} if $\cG'$ is of the form (\ref{actionofdiffeomorphisms}) for  some local diffeomorphism $\wh \varphi$ of $M$ with $\wh \varphi(M_o) \subset M_o$.
%\end{itemize}
%\end{definition}
%\par
%Diffeomorphic (super) gravities are necessarily also  weakly diffeomorphic, but it is reasonable to expect that, without any additional constraint,  the converse is not true. \par
%\smallskip
%One could certainly conceive a sort of ``rheonomic principle'', which resembles the principle stated  in   \cite{CDF},  according to which  ``the only allowed   theories of (super) gravity are those 
%for which  solutions to constraints and equations are weakly diffeomorphic if and only if they are strongly diffeomorphic".  A principle of this kind could be a useful guiding rule 
%for  constructions of (super) gravity theories,  but for the moment this  topic is  out of our scope. \par
This suggests  the following two notions:
\par
\bigskip
%\vskip 0.5cm
%\centerline{\it Generalized Principle of   Infinitesimal General  Covariance}
%\smallskip
\moveleft 0.6cm
\vbox{\begin{itemize}
\item[]
  {\it A collection $\cE_o$ of constraints and equations  on the physical fields of  (super) gravities of type $\gg$ satisfies the  {\rm Generalized Principle of  Infinitesimal General  Covariance} if:
\begin{itemize}
\item[i)] there exists a system $\cE$ of constraints and equations on $(\cD,g,\n)$, so that any (local) solution of $\cE$ determines physical fields which solve $\cE_o$ and every (local) solution of $\cE_o$ is of this form;
\item[ii)] the class of (local) solutions of $\cE_o$ is invariant under all actions \eqref{actionofdiffeomorphisms}, where $\cG$ is given by a solution of $\cE$ and
$\varphi$ is of the form $\varphi=\Phi^X_t$ for some $X \in \gX_{loc} (M)$.
\end{itemize}
}
\item[]
\item[]  {\it The system  $\cE_o$ is said {\rm manifestly covariant} if there exist a system $\cE$ as in (i) which is of tensorial type.
%\begin{itemize}
%\item[a)] $\cE_o$ is given by restrictions to  $TM|_{M_o} = TM_o + \cS$ of 
%a system $\cE$ of constraints and equations on $M$   of tensorial type; 
%\item[b)] any solution of $\cE_o$ can be locally obtained as restriction to $TM|_{M_o} = TM_o + \cS$ of some 
%solution of $\cE$.
%\end{itemize}
}
\item[]
\end{itemize}}
\par
Any manifestly covariant  system $\cE_o$  {\it automatically satisfies (ii)} and hence {\it also the Generalized
Principle of General Covariance}. \par
\medskip
Now, for a given (super) gravity $\cG= ((M, M_o, \cD), (g, \n))$ of type $\gg$, let us consider the following class of (local) vector fields on $M$  
%\beq \gX_{loc} (M; M_o) \cup  \gX_{loc} (M; \cS) \subsetneq \gX(M)\qquad \text{where}\eeq
%\par
%\vskip -0.4cm
$$\gX_{loc} (M; M_o) = \{X \in \gX_{loc}(M):X_{x} \in T_x M_o, \ x \in M_o\}\ ,$$
$$\gX_{loc} (M; \cS) = \{X \in \gX_{loc}(M):X \in \Gamma_{loc}(\cD),  \ (\n_s X)_x = 0, \  s \in  \cD_x,\ x \in M_o\}\ .$$
%$$\gX_{loc} (M; \cS) = \{\ X \in \gX_{loc}(M)\ : \ X_{x} \in \cS = \cD|_{M_o}\ \text{for any} \ x \in M_o\ \}\ .$$
%The classes $\gX_{loc} (M; M_o) $ and $\gX_{loc} (M; \cS)$ can be considered as   ``gauged versions''  of the subspaces $V$ and $S$  of   $ \gg = \so(V) +  V + S$. \par
Clearly, any  $X \in \gX_{loc}(M_o)$ admits an extension $\wh X \in \gX_{loc}(M; M_o) $ and one can check that
any local section $s$ of  $ \cS$ admits   an extension $\wh s \in \gX_{loc}(M; \cS) $.
\par
The actions of the fields in $\gX_{loc}(M; M_o) $  can 
 be considered as  generalizations of   the actions of the vector fields of   $M_o$. In fact,  for any $X \in \gX_{loc}(M_o)$
 with extension $\wh X \in  \gX_{loc}(M_o) $,   the family of metrics  $ \Phi^{X}_t{}_* (\wh g)$ coincides with  the family of gravitons $\wh g^{t}$  of the  (super) gravities   $\cG_t = \Phi^{\wh X}_t{}_*(\cG)$. \par
As we will see later (\S \ref{representing}),  the actions   of  fields in $\gX_{loc} (M; \cS)$ coincide  with the supersymmetries of simple $4D$-supergravity and 
other supergravities.\par 
\bigskip
In other words, 
those supergravity theories are invariant under the class of vector fields 
\beq \label{algebretta} \gX_{loc} (M; M_o) + \gX_{loc} (M; \cS)\ ,\eeq
which is properly included in $\gX_{loc}(M)$. \par
 %More precisely, if  for any $X \in \gX_{loc}(M)$ we denote by  $\d_X$  the map, which   associates to the triple $(\wh g, \vartheta, A)$ the corresponding variations
%$$(\wh g, \vartheta, \bA) \overset{\d_X}\longmapsto  (\d_X \wh g, \d_X \vartheta, \d_X \bA)\ ,$$
%then  for any   $X, X' \in \gX_{loc}(M)$ and $\l, \m \in \bR$, 
% \beq \label{vectorspace} \l \d_X + \m \d_{X'} = \d_{\l X + \m X'}\quad  \text{and}\quad  \d_X \circ \d_{X'} - \d_{X'} \circ \d_{X}  = \d_{[X, X']}\eeq
%i.e. $\gS\=\{\d_X, \ X \in  \gX_{loc}(M)\}$ {\it has a structure of Lie (pseudo-)algebra\/}. \par
%Notice that this claim is  only a very direct consequence of the relation (forced by our definitions) between the variations $\d_X$  and Lie derivatives along vector fields, but it is very useful nonetheless. \par
%\par
%These two classes and, as we will see in \S \ref{firstactions},   their elements  act on the physical fields of (super) gravities as many transformation rules  of \par
%
%and hence  equations on gravitons of a theory satisfying   the generalized Principle of Infinitesimal General Covariance satisfy  also the  Principle of Infinitesimal General Covariance as classically understood.\par
%\medskip
%\smallskip
%\noindent
\medskip
We    recall that 
a class $\cA \subset \gX_{loc}(M)$ of   vector fields 
 is called
{\it Lie pseudo-algebra\/}  if for any $\l,\mu \in \bR$   and any pair $X, X' \in \cA$, defined on  two open subsets $\cU, \cU' \subset M$,   
the fields $\l X + \mu X'$ and $[X,X']$ on $ \cU \cap \cU'$ 
are  both elements of   $\cA$. Lie pseudo-algebras share many basic properties with  usual Lie algebras of vector fields  (see e.g.  \cite{SS}).
\par
\medskip
It is hardly to be expected that  brackets between elements in  \eqref{algebretta}  are still  in \eqref{algebretta}, i.e. that  \eqref{algebretta} is a Lie pseudo-algebra.  Hence, if one is looking for  a Lie pseudo-algebra of symmetries, it is more natural to consider   the whole   $\gX_{loc}(M)$. 
In fact,  as we will shortly see,   equations of simple $4D$-supergravity are manifestly covariant and hence invariant under all elements in  $\gX_{loc}(M)$.
%
%\vskip 2 cm
%
%This 
%might be one reason for the fact that the space of symmetries of certain supergravities is not closed under commutators. \par
% On the other hand, whenever in a supergravity theory one is able to show that  the  transformations rules for graviton, gravitino and auxiliary field coincide with  those given in 
%(\S \ref{Levi-Civita supergravities and the transformations laws for their physical  fields}), one is  entitled to   identify those fields 
%with the graviton, gravitino and A-field considered in this paper and the fact that $\gS$ is a Lie pseudo-algebra represents an  important  property  of the theory, which is usually 
%proved only through tedious  computations   in coordinates. 
\section{Levi-Civita supergravities and the transformations laws for their physical  fields}
\label{Levi-Civita supergravities and the transformations laws for their physical  fields}
\subsection{Levi-Civita supergravities}
Let  $\cG = ((M, M_o, \cD), (g,  \n) )$ be a  (super) gravity of type $\gg$. 
At any  $x \in M$,   the torsion $ T_x$ of $\n$  decomposes  into a sum of the form  
\beq \label{decoT}  T_x = T^{\cD^\perp}_x + T^{\cD}_x + \cC_x^{\cD, \cD^\perp; \cD} + \cC_x^{\cD, \cD^\perp; \cD^\perp}
 +  \cH_x^{\Lambda^2\cD^\perp; \cD} +  \cH_x^{\Lambda^2\cD; \cD^\perp}\ , \eeq
with  summands belonging to the following $\so(\cD^\perp_x)$-modules:
$$T^{\cD^\perp}_x   \in \Hom(\cD^\perp_x\wedge \cD^\perp_x, \cD^\perp_x)\ ,\quad T^{\cD}_x   \in \Hom(\cD_x\wedge \cD_x, \cD_x)\ ,$$
$$\cC^{\cD, \cD^\perp; \cD}_x   \in \Hom(\cD_x\times \cD^\perp_x, \cD_x) \ ,\qquad
\cC^{\cD, \cD^\perp; \cD^\perp}_x  \in \Hom(\cD_x\times \cD^\perp_x, \cD^\perp_x) 
\ , $$
$$\cH_x^{\Lambda^2\cD^\perp; \cD} \in \Hom(\cD^\perp_x \wedge \cD^\perp_x, \cD_x)\ ,\qquad \cH_x^{\Lambda^2\cD; \cD^\perp} \in \Hom(\cD_x \wedge \cD_x, \cD^\perp_x)\ .$$
\par
\smallskip
\noindent Notice   that {\it the decomposition \eqref{decoT} is preserved by any action \eqref{actionofdiffeomorphisms}\/}. 
\par
\medskip
Since $\n $ preserves $\cD$ and $\cD^\perp$,  it follows that   
\beq \label{nonvanishing} \cH_x^{\Lambda^2\cD; \cD^\perp} = - \cL^g_x 
\simeq - \cL^\gg_o\ .
\ \eeq
From this, at any $x_o \in M_o$ and
for any $X, Y \in \gX(M_o)$ with $[X, Y]|_{x_o} = 0$,   the value  $\left.T^D_{X  Y}\right|_{x_o}$ of the torsion of the metric connection $D$ is  equal   to 
$$\left.T^D_{XY}\right|_{x_o} = \left.(\left.\pi^{\cD^\perp} \right|_{TM_o})^{-1}\left(\n_X( \pi^{\cD^\perp}(Y)) - \n_Y ( \pi^{\cD^\perp}(X)) \right)\right|_{x_o}= $$
$$ =\left.(\left.\pi^{\cD^\perp} \right|_{TM_o})^{-1}\left(\pi^{\cD^\perp} (T_{XY}) \right)\right|_{x_o}=  (\left.\pi^{\cD^\perp} \right|_{TM_o})^{-1}\circ$$
{\small \beq
\label{torsionmetric}
\circ \left.\left(T^{\cD^\perp}_{X^\perp Y^\perp} + \cC^{\cD, \cD^\perp; \cD^\perp}(\vartheta(X), Y^\perp) - \cC^{\cD, \cD^\perp; \cD^\perp}(\vartheta(Y), X^\perp)
-\cL^g(\vartheta(X), \vartheta(Y))\right)\right|_{x_o}
\eeq}
(here $X^\perp$, $Y^\perp$ denote the components of $X$,  $Y$ along $\cD^\perp$).
Hence,  for any admissible  extended (or super) Poincar\`e algebra  $\gg = \so(V) + V + S$, the torsion $T^D$ has a non trivial term,  depending quadratically on  $\vartheta$. Due to this,  {\it there is no way to require the  vanishing of $T^D$ for all values of $\vartheta$}, in contrast with the well-known property  of   Levi Civita connections.\par
 However the following theorem holds (in the statement,   $\Sym(\cD^\perp_x) $ denotes the space of  endomorphisms of $\cD^\perp_x$ that are symmetric w.r.t. $g_x$).   \par
\medskip
\begin{theo}\label{existenceLC} For any  $((M, M_o, \cD), g)$ that  satisfies Definition \ref{gravityfields2} (i),  there exists a  connection $\n$ satisfying also (ii),  (iii)  and the constraints  
\beq \label{defLC} 
T^{\cD^\perp}_x=0\ \ \text{and}\ \ \ \cC^{\cD, \cD^\perp; \cD^\perp}_x\in\cD^*_x\otimes \Sym(\cD^\perp_x)\ \ \  \text{at any} \ x \in M\ .
\eeq
This connection is uniquely determined up to the field $C$, defined  in \eqref{tensorC}. In particular, $\n$ is  unique whenever $\go(S, \b)\cap C_{\ggl(S)}(\cC \ell(V)) = 0$.
 \end{theo}
\begin{pf}
Assume that $g$ is an extended metric on $(M, M_o, \cD)$ satisfying Definition \ref{gravityfields2}.  For a fixed choice of local frames in $\operatorname{O}_g(M, \cD)$,  let  
 $\n$ be  the unique,  locally defined, $\cD$ and $\cD^\perp$ preserving  connection, for which the field $C$  in \eqref{tensorC} is $0$ and  for any  $v,w,z\in \cD^\perp$, $s\in \cD$
 \beq
g(\n_v w,z)=\frac{1}{2}\left(v \cdot g(w,z)+w\cdot g(z,v)- z\cdot g(v,w)+\phantom{a\pi^{\cD^\perp}aaaaaaaaaaaaa}\right.$$
$$ \phantom{aaaaaaa}+ \left.g(\pi^{\cD^\perp}([v,w]), z)- g(\pi^{\cD^\perp}([w,z]),v)-g(\pi^{\cD^\perp}([v, z]),w)\right)\ ,
\eeq
\beq
g(\n_ s w,z)=\frac{1}{2}\left( s \cdot g(w,z )+g(\pi^{\cD^\perp}([s, w]), z)-g(\pi^{\cD^\perp}([s,z]), w)\right)\ .
\eeq
\par
One can check  that it satisfies \eqref{defLC} and (ii), (iii). Using a partition of unity, one gets  the existence part of the theorem.\par
\smallskip
 About
 the uniqueness part, assume that 
$\cG = ((M, M_o, \cD), (g, \n))$ and $\wt \cG =((M, M_o, \cD),  (g,  \wt \n))$  are two (super) gravities of the same  type $\gg$, both satisfying \eqref{defLC}.  
Fix a local  identification \eqref{identification} so that we may consider the decompositions $\n = \n^o + C$ and $\wt \n = \wt\n^o + \wt C$ described  in \eqref{tensorC}.
By definitions,   for any $X \in \gX(M)$, the operators $\n^o_X$ and $\wt \n^o_X$  act on the vector  fields in $\cD^\perp$ just as  the covariant derivations $\n_X$ and $\wt \n_X$, while they act on the fields in  $\cD$ 
by means of the corresponding spinorial  connections. In particular,  $\n^o$ and $\wt \n^o$ are uniquely determined by  their restrictions $\n^o|_{\gX(M) \times \cD^\perp}$ and $\wt\n^o|_{\gX(M) \times \cD^\perp}$. \par
On the other hand, by definitions, 
\beq \label{preS1} \wt \n^o|_{\gX(M) \times \cD^\perp} = \n^o|_{\gX(M) \times \cD^\perp} + F\ ,\eeq
for some suitable tensor field  $F$  taking values  in  $T^* M \otimes\so(\cD^\perp)
$, i.e. so that for any $X \in \gX(M)$, $v, v' \in \G(\cD^\perp)$ 
\beq \label{S1} g(F_{X}(v), v') + g(v, F_{X}(v')) = 0\ . \eeq
% Conversely,  for any  $\wt \n = \n + C$ satisfying satisfying (ii) - (iv) of  Definition \ref{gravityfields2}  and a tensor field $F $ in $T^* M \otimes\so(\cD^\perp)
%$, the  connection $\wt \n' =\n'+C$,  with $\n'$ as in   \eqref{preS1},  also satisfies (ii) -  (iv) of Definition \ref{gravityfields2}.\par
%Notice also that, by definition of $C$, for any connection $ \n' $ of a (super) gravity,  the torsion $ T'$ is so that $\pi^{\cD^{\perp}} \circ  T$ is independent on $C$. Hence  $\pi^{\cD^{\perp}} \circ  T'$ is uniquely determined by the torsion $ T$ of $ \n$ and the tensor field $F$.\par
Now, for a given  $F$ in $T^* M \otimes\so(\cD^\perp)
$,  let us denote by 
$\partial_FT^o=T^o- \wt T^o$ 
the difference between the torsions $T^o$ and $\wt T^o$ of the connections $\n^o$  and $\wt \n^o$, respectively. Simple arguments based just on definitions imply that
$$
\partial_FT^o_{XY} =F_X(Y)-F_Y(X)\ ,\qquad \text{for any}\ X, Y \in \gX(M)
$$
and that,  for any $x \in M$, the map
$$\varphi_1: \cD^*_x \otimes \so(\cD^\perp_x)  \longrightarrow  \cD^*_x \otimes \End(\cD^\perp_x) \ ,\phantom{aaaaaaaaaaaaaaaaaaaaaaaaaaaa}$$ 
\beq  \label{corre1} \phantom{aaaaaaaaaaaaaaaaaaaaaa} \varphi_1\left( F_x|_{\cD_x \times \cD^\perp_x}\right) \= \pi^{\cD^\perp} \circ  (\partial_F T^o)|_{\cD_x \times \cD^\perp_x}\eeq
coincides with the trivial embedding of $\cD^*_x \otimes \so(\cD^\perp_x)$ into $ \cD^*_x \otimes \End(\cD^\perp_x) $. 
Due to this,  by the fact  that the  antisymmetric parts of the tensors  $\cC^{\cD, \cD^\perp; \cD^\perp}_x$  of $\n$ and $\wt \n$ are both  $0$, one gets    $F_x|_{\cD_x \times \cD^\perp_x} = 0$ at any $x \in M$. \par
Consider now the map 
$$\varphi_2: \cD^{\perp*}_x \otimes \so(\cD^\perp_x)  \longrightarrow \Lambda^2 \cD^{\perp*}_x \otimes \cD^{\perp}_x \ ,\phantom{aaaaaaaaaaaaaaaaaaaaaaaaaaaa}$$ 
\beq \label{corre}  \phantom{aaaaaaaaaaaaaaaaaaaaaa}\varphi_2\left(F_x|_{\cD^\perp_x \times \cD^\perp_x} \right) =  \pi^{\cD^\perp} \circ (\partial_F T^o)|_{\cD^\perp_x \times \cD^\perp_x}  \ .\eeq
We claim that this is  a vector space  isomorphism between  $\cD^{\perp*}_x\otimes \so(\cD^\perp_x)$ and $\Lambda^2 \cD^{\perp*}_x \otimes \cD^\perp_x$. In fact, if we identify $T_x M$ with  $V + S$ by means of a frame in $\operatorname{O}_g(M, \cD)$, the map
 $F_x|_{\cD^\perp_x \times \cD^\perp_x}$ is identifiable with  an element of 
$$V^* \otimes \so(V, <,>) \simeq V^* \otimes \Lambda^2 V^*\ ,$$ while  $\pi^{\cD^\perp} \circ (\partial_F T^o)|_{\cD^\perp_x \times \cD^\perp_x} $ is identifiable with  an element of $\Lambda^2 V^* \otimes V  \simeq \Lambda^2 V^* \otimes V^*$.  The map \eqref{corre} is equal to the  so-called ``Spencer operator''
$$ \partial: V^* \otimes \Lambda^2 V^* \longrightarrow   \Lambda^2 V^* \otimes V^*\ ,$$
\beq 
(\partial \a)(v_1, v_2, w) = \a(v_1, v_2, w) - \a(v_2, v_1, w)\ ,\eeq
which is well known to be an isomorphism. Due to this,   since    $\n$ and $\wt \n$  has  $T^{\cD^\perp}\equiv 0$, then 
 $F_x|_{\cD_x^\perp \times \cD_x^\perp} = 0$ 
at  any  $x\in M$. \par
Hence, $F \equiv 0$ and $\n^o = \wt \n^o$. 
The claim is then a  consequence of   the fact that variations  of  $C$ do  not affect  $T^{\cD^\perp}$ and $\cC^{\cD, \cD^\perp; \cD^\perp}$. 
\end{pf} 
This result motivates the following definition.\par
\begin{definition} A (super) gravity $((M, M_o, \cD), (g, \n))$  is called 
{\it Levi-Civita} if   $\n$ satisfy \eqref{defLC}.
%(\footnote{It should be noticed that Theorem \ref{existenceLC} does not depend on the submanifold $M_o$ and it could have been formulated purely in terms of the triple $(M,\cD,g)$. From this point of view, the connection  $\n$ satisfying \eqref{defLC} could be regarded as an higher-codimension analogue of the Tanaka-Webster connection in CR geometry (\cite{Tan, Web}).}).
In this case,  $\n$ is called a {\it Levi-Civita connection  of $g$}.
\end{definition}
Let  $ \cG = ((M, M_o, \cD), (g, \n))$ be a Levi-Civita (super) gravity. By  \eqref{torsionmetric}, the value of the torsion  of the metric connection $D$ on commuting fields  is 
$$ T^D_{XY} = (\left.\pi^{\cD^\perp} \right|_{TM_o})^{-1}\circ$$
$$
\circ \left.\left( \cC^{\cD, \cD^\perp; \cD^\perp}(\vartheta(X), Y^\perp) - \cC^{\cD, \cD^\perp; \cD^\perp}(\vartheta(Y), X^\perp)
-\cL^g(\vartheta(X), \vartheta(Y))\right)\right|_{M_o}\ 
.$$
This shows that $T^D$ (and hence $D$, being any metric connection  recoverable  from  its torsion, through the  associated contorsion) is completely determined by 
the graviton, the gravitino and the tensor field $\cC^{\cD, \cD^\perp; \cD^\perp}|_{M_o}$. \par
\bigskip
 A common assumption  in supergravity  is $\cC^{\cD, \cD^\perp; \cD^\perp}|_{M_o}Ê= 0$ (see \S \ref{representing}).  It is therefore convenient to introduce the following definition. 
% A condition satisfying the Principle of Infinitesimal General Covariance  and implies that  $\cC \equiv 0$  is the following. \par
 \begin{definition} A (super) gravity $\cG = ((M, M_o, \cD), (g, \n))$  is called 
{\it strict Levi-Civita} if the torsion of $\n$ satisfies the conditions
$T^{\cD^\perp} \equiv 0 \equiv \cC^{\cD, \cD^\perp; \cD^\perp}$.
\end{definition}
\medskip
%Recalling that   any metric covariant derivation  is completely determined by its torsion, we see that   in a strict Levi-Civita (super) gravity the metric connection 
%is completely determined by the graviton and gravitino. 
Since the difference between spinor and metric connections is  given by the A-field (and the field $C$ in \eqref{tensorC}, 
in special signatures), 
it follows that   {\it all physical fields of strict Levi-Civita (super) gravities
are completely determined by the graviton, gravitino and A-field} (and, sometimes,  by  $C$).  
\subsection{Transformations rules for gravitons,  gravitinos and $A$-fields}
\label{firstactions}
In this section we give explicit formulae for the  actions of vector fields in $\gX_{loc}(M; \cS)$ on the graviton, gravitino and $A$-field of a strict Levi-Civita (super) gravity. We  perform computations in local coordinates and components
to show that the obtained expressions nicely match the well-known rules of simple 4D-supergravity and other supergravities.
\par
\smallskip
%The corresponding actions   on the $A$-field  (and the $B$-field) will be given in the next section.\par
%\smallskip
Let $\cG = ((M, M_o, \cD), (g, \n))$ be a {\it strict Levi-Civita supergravity}  and $(E_a, E_\a )$ a (local) field of  $g$-orthonormal frames  with $ E_a \in \cD^\perp$,  $E_\a \in \cD$ and w.r.t. which the Levi form $\cL^g$ has constant components $\cL^a_{\a \b}$.  
%
%
%$\imath^{(\cdot)}$  a   of .  Given an orthonormal  basis 
%$(E_a^o, E^o_\a)$ for $(V + S, (\cdot, \cdot))$, with vectors $E_a^o$ in $V$ and vectors $E_\a^o$ in $S$. Then the   family  of isomorphisms  $\imath^{(x)} \= \s_x: V + S \longrightarrow T_x M$
%determine a field of vielbeins on $M$
%$$(\ E_a \= \imath^{(x)}(E^o_a)\ ,\quad  E_\a \= \imath^{(x)}(E^o_\a)\ )$$
%with $ E_\a|_x \in \cD^\perp_x$,  $E_\a|_x \in \cD_x$ for any  $x \in \cU$. 
Let also  $(E^a, E^\a)$ be the dual coframe field.\par
\medskip
Now, if we consider the $\wh g$-orthonormal coframes  $(e^a =  \left.E^a\right|_{T M_o})$ on $M_o$,  we have  that the graviton, the gravitino and  the A-field 
are of the form 
\beq
\label{gravitonegravitino}
\wh g =   \h_{ab}\  e^a \otimes e^b\ , \ \ \  \vartheta =  \psi^\a_b \ E_\a|_{M_o} \otimes e^b\ , \ \  \bA = \Gamd \a a \b \ E_\a|_{M_o} \otimes e^a \otimes E^\b|_{\cS} \ ,\eeq
where $\h_{ab}  = \epsilon_a \d_{ab}$,  and  $\psi^\a_b$, $\Gamd \a a \b$ are suitable smooth functions. Indeed,  $\wh g$, $\vartheta$, $\bA$  are  the restriction to $TM_o$ and $\cS$ of the tensor fields of $M$
$$ g (\pi^{\cD^\perp}(\cdot), \pi^{\cD^\perp}(\cdot)) = \h_{ab}ÊE^a \otimes E^b \ ,\qquad  \pi^\cD = E_\a \otimes E^\a\ ,$$
$$ - \pi^\cD \circ T = - T_{ab}^\a\  E_\a \otimes E^a \otimes E^b - T_{a\b}^\a\  E_\a\otimes E^a \otimes E^\b - T_{\g\b}^\a\  E_\a\otimes E^\g \otimes E^\b\ .$$
In particular,  the $\psi^\a_b$ are the components of the 1-forms  $E^\a|_{TM_o} = \psi^\a_b e^b$, while the $\Gamd \a a \b$ are the functions
$\Gamd \a a \b= - T_{a\b}^\a|_{M_o} - \psi^\gamma_a\cdot T_{\g\b}^\a|_{M_o}$.\par
%
%
%
%The graviton $\wh g$ is completely determined by the coframes 
%$$(e^a =  \left.E^a\right|_{T M_o})\ ,$$
%which are   orthonormal w.r.t to   $\wh g$.  Similarly, if we consider   the tensor field $\pi^\cD = E_\a \otimes E^\a$, which 
%at any $x \in M$  projects orthogonally any vector  onto the subspace $\cD_x \subset T_x M$, the gravitino and the A-field are determined as 
%$$\vartheta = \left.\pi^\cD\right|_{TM_o}\ , \qquad \bA = \left.( -   \pi^\cD \circ T )\right|_{ T M_o \times \cS}\ .$$
%  In particular, 
%   the graviton, gravitino the A-field are of the form 
%\beq
%\label{gravitonegravitino}
%\wh g =   \h_{ab}\  e^a \otimes e^b\ , \ \ \  \vartheta =  \psi^\a_b \ E_\a|_{M_o} \otimes e^b\ , \ \  \bA = \Gamd \a a \b\ E_\a|_{M_o} \otimes e^a \otimes E^\b|_{\cS} \eeq
%where $\Gamd \a a \b=A_{a\b}^\a|_{M_o}+\psi^\gamma_a\cdot A_{\g\b}^\a|_{M_o}$.
%Finally, we denote by $(e_a)$ the field of $\wh g$-orthonormal frames on $M_o$ uniquely determined by 
%$Ê \pi^{\cD^\perp}(e_a)=E_a|_{M_o}$.
\par
\bigskip
Motivated by the above remark, we call  {\it variations of the graviton, the gravitino and the A-field along $X \in \gX_{loc}(M)$}  the fields on $M_o$ defined by  
$$ \d_X e^a:= (\cL_X E^a)|_{TM_o}\ ,\ \d_X \vartheta:=(\cL_X \pi^\cD)|_{TM_o}\ ,\ \d_X \bA:=  -  (\cL_X (\pi^\cD \circ T))|_{T M_o \times \cS}$$
where  ``$\cL_X$'' denotes the usual Lie derivative along the vector field $X$. 
%We recall that, according to our previous definition, the set of solutions to the constraints and equations of  theory of (super) gravity must be  invariant under
%these infinitesimal variations, which  are therefore {\it symmetries of the theory}. 
As we will see in \S \ref{representing}, the variations along the vector fields in $\gX_{loc} (M; \cS)$ correspond to the so-called  ``supersymmetry transformations''  in simple $4D$-supergravity  and other supergravity theories. We thus consider the following definition.\par
%To make this correspondence more precise, it is convenient to consider the following definition. \par
% 
% Using just definitions, one has that  
%{\it the gravitino  $\vartheta$ is at all points of the form $\vartheta_x = \psi^\a_m(x) E_\a \otimes e^m$, where the components $\psi^\a_m(x)$ 
%constitute an object with  the standard  properties of  spin-$\frac{3}{2}$ gravitinos} (see e.g. \cite{WB}, Ch. XVII). \par
%
%
%
%Given  
%i.e.   a local admissible frame, that is a field of vielbeins $(E_i, E_\a)$ of $(M, g)$, adapted to the Levi form $\cL$, with $E_i\in\cD^\perp$ and $E_\a\in\cD$. We denote the dual coframe by $(E^i, E^\a)$ and the corresponding fields on $M_o$ by 
%%$E_i|_{M_o}, E_\alpha|_{M_o}, E^i|_{M_o}, E^\alpha|_{M_o}$ their restrictions to the body $M_0$.
%\be
%\label{riferimentibody}
%\wh e^i = \imath^* E^i\ ,\qquad \vartheta^\a = \imath^* E^\a\ ,
%\ee
%so that $\cL$ is uniquely determined by the constants
%$\cL^i_{\a \b} \= E^i( \cL^g(E_\a ,E_\b))$.
%The graviton and the gravitino are (the restrictions to $T_x M_o$ of) the fields 
%\be
%\label{gravitonegravitino}
%\wh g = \sum_i \epsilon_i\ \wh e^i \otimes \wh e^i\ ,\qquad \vartheta = \sum_\a \vartheta^\a \otimes E_\a|_{M_o}\ .
%\ee
%The associated field of vielbeins $(\wh e_j)$ for $(M_o,\wh g)$ is uniquely determined by 
%\beq
%\label{ej}
%  \pi^{\cD^\perp}(\wh e_j)= \wh e_j - \vartheta(\wh e_j)=\wh e_j -  \vartheta^\a(\wh e_j)E_\a|_{M_o}=E_j|_{M_o}\ .\eeq
%\par
\begin{definition}
\label{spinorone}
Let $\varepsilon = \varepsilon^\a E_\a|_{M_o}$ be  a (locally defined) spinor field in $\cS$.  We call ({\it super}) {\it variations along $\varepsilon$}   the infinitesimal variations
$$\d_{\varepsilon} e^a \= \d_{X^{(\varepsilon)}} e^a\ ,\qquad  \d_{\varepsilon} \vartheta\= \d_{X^{(\varepsilon)}} \vartheta\ , \qquad  \d_{ \varepsilon} \bA\= \d_{X^{(\varepsilon)}} \bA\ ,$$ 
determined by  an arbitrary vector field   $X^{(\varepsilon)} = \gX_{loc}(M;\cS)$ with  $X^{(\varepsilon)}|_{M_o} = \varepsilon$.
%\beq
%\label{spinorone}\left. X^{(\varepsilon)}\right|_{M_o} =\varepsilon^\a \left.E_\a\right|_{M_o}\ ,\qquad 
%\left.\wt \n_{E_\alpha}X^{(\varepsilon)}\right|_{M_o}=0\ ,\ \ 1 \leq \a \leq n^S\ .
%\eeq
\end{definition}
The (super) variations along $\varepsilon$ are clearly  determined by the  functions   $\d_\varepsilon e^a_b$,  $\d_\varepsilon \psi^\a_a$, $ \d_{\varepsilon}\psi^b_a$, $\d_\varepsilon \Gamd \a a \b$ and $ \d_\varepsilon \Gamd b a \b$ defined by the relations
%
%$(e^a, E^\a|_{\cS})$ and the frames $(E_a|_{M_o}, E_\a|_{M_o})$ (which  are $g$-orthonormal, but {\it not} dual to $(e^a, E^\a|_{M_o})$), defined by 
\beq 
\label{funzioncineI}
\d_{\varepsilon} e^a:= (\d_\varepsilon e^a_b)\   e^b\ ,\quad \d_{\varepsilon} \vartheta:= (\d_\varepsilon \psi^\a_a)\ E_\a|_{M_o}\otimes e^a + (\d_{\varepsilon}\psi^b_a)\ E_b|_{M_o} \otimes e^a\ ,
\eeq
\beq
\label{funzioncineII} \d_{ \varepsilon} \bA := (\d_\varepsilon \Gamd \a a \b) \ E_\a|_{M_o} \otimes e^a \otimes E^\b|_{\cS} +  (\d_\varepsilon \Gamd b a \b)\  E_b|_{M_o} \otimes e^a \otimes E^\b|_{\cS}\ .
\eeq
We now compute explicitly   those functions, proving also that 
they are independent of the choice of the extension $X^{(\varepsilon)}$. In the following, $(e_a)$ is the $\wh g$-orthonormal frames field on $M_o$ defined by $\pi^{\cD^\perp}(e_a) = E_a|_{M_o}$.\par
%that satisfies the conditions \eqref{spinorone}.  \par
\begin{prop} \label{transformations1} Given a   spinor field $\varepsilon = \varepsilon^\a E_\a|_{M_o}$,  the components of the corresponding (super) variations  of graviton and gravitino are of the form
\beq \label{variationgraviton} \d_\varepsilon e^b_a=  -  \varepsilon^\a \psi^\b_a \cL^b_{\a \b} + L^b_a \quad  \text{for some}\ L = (L^b_a) \in  \so(V)\ , \eeq
\beq \label{variationgravitino} \d_\varepsilon \psi^\a_a=   e_a( \varepsilon^\a) + \varepsilon^\b (\GamD \a a \b + \Gamd \a a \b+\psi_{a}^{\gamma}\bB_{\b\gamma}^{\alpha})\ , \eeq
\beq \label{variationgravitino1}  \d_\varepsilon \psi^b_a = \varepsilon^\a \cL^b_{\a \b} \psi^\b_a \ ,\eeq
where
%\begin{itemize}
%\item[--] $\cL^a_{\a \b}$ are the components of the Levi form w.r.t. $(E_a, E_\a)$; 
%\item[--] 
$\GamD \a a \b$,  $\bB_{\beta\gamma}^{\alpha}$ are  the  Christoffel symbols   $\GamD \a a \b \= E^\a(\n_{e_a} E_\b)$ and the components of the $\bB$-field $\bB_{\beta\gamma}^{\alpha}\=E^{\alpha}(\bB_{E_\beta E_\gamma})$, respectively.
%\end{itemize}
\end{prop}
\begin{pfns} 
For simplicity of notation, let us  denote by the symbol  `` $\varepsilon$'' also  the  field $X^{(\varepsilon)} \in \gX_{loc}(M; \cS)$. We first need one simple observation. In order to compute the functions which determine the variations \eqref{funzioncineI}, one has to evaluate the tensor fields $\cL_\varepsilon E^a$, $\cL_\varepsilon \pi^\cD$ on elements of $TM_o$. 
One can also check that it is always possible to extend the field of $\wh g$-orthonormal frames $(e_a)$ on $M_o$ to vector fields  $(e_a)$  on an open subset $\cU \subset M$ so that $\pi^{\cD^\perp}(e_a)=E_a$  and 
%\pi^{\cD^\perp}(\wh e_j)=E_j
%$$
%and 
%$$ \n_{E_\a} \pi^{\cD}(\wh e_j)|_{M_o} = 0 $$
%for any $\a = 1, \dots, n^S$. \par
$\n_{E_\a} \pi^{\cD}(e_a)|_{M_o} = 0$. By tensoriality, one is allowed to evaluate $\cL_\varepsilon E^a$ and $\cL_\varepsilon \pi^\cD$ on these special extensions and then restrict the result to $M_o$.

\par% for any $1 \leq a \leq n$, $1 \leq \a\leq n^S$ at all points of $\cU$.\par
\smallskip
With these remarks in mind, one has  that 
$$\d_\varepsilon e^b_a =(\cL_\varepsilon E^b)(e_a) = - E^b([\varepsilon, e_a])  =  - E^b(\n_\varepsilon e_a)  + E^b(\n_{e_a} \varepsilon) 
+ $$
$$ + E^c(e_a) E^b(\cC^{\cD, \cD^\perp; \cD^\perp}(\varepsilon, E_c)) 
 + E^\a(e_a) E^b(\cH^{\L^2 \cD; \cD^\perp}(\varepsilon, E_\a))$$
and that  the matrix  $L^b_a = - E^b(\n_\varepsilon e_a) |_x$ belongs to $\so(V)$ for any $x \in M_o$.  From
 $E^b(\n_{e_a} \varepsilon)=0$, $\cC^{\cD, \cD^\perp; \cD^\perp}= 0$ and  \eqref{nonvanishing}, equality  \eqref{variationgraviton} follows. \par
Similarly, we have that 
$$\d_\varepsilon \psi^\a_a=\left.E^\a( (\cL_\ve \pi^\cD)(e_a))\right|_{M_o}
= \left.E^\a\left([\varepsilon,  \pi^\cD( e_a)]\right)\right|_{M_o} - 
\left.E^\a\left(\pi^\cD([\varepsilon,  e_a])\right)\right|_{M_o}\ . $$
\smallskip
Since
$$E^\a([\varepsilon,  \pi^\cD( e_a)]) = E^\a(\n_{\varepsilon} \pi^\cD( e_a)) - E^\a(\n_{ \pi^\cD( e_a) } \varepsilon) - E^\a(T^\cD_{ \varepsilon\ \pi^\cD( e_a)})\ ,$$
$$E^\a\left(\pi^\cD([\varepsilon,  e_a]\right) = E^\a\left([\varepsilon,  e_a]\right) = E^\a(\n_{\varepsilon}  e_a) - E^\a( \n_{e_a} \varepsilon) - E^\a(T_{ \varepsilon e_a}) = $$
$$ = E^\a(\n_{\varepsilon}  \pi^\cD(e_a)) - E^\a(\n_{e_a} \varepsilon) - E^\a(\pi^\cD(T_{ \varepsilon e_a}))\ ,$$
we have that 
%Hence, from \eqref{spinorone}, the hypothesis on the   fields $e_a$  and the definitions of gravitino and spinor connection, we get  that 
$$\d_\varepsilon \psi^\a_a  =E^{\a}(\n_{e_a}\ve+ \bA_{e_a \varepsilon}+\bB_{\varepsilon\vartheta(e_a)})=E^\a(\bD_{e_a} \varepsilon + \bB_{\varepsilon\vartheta(e_a)})=
$$
$$=e_a (\varepsilon^\a)+ \varepsilon^\b E^\a(\bD_{e_a} E_\b) + E^\a(\bB_{\varepsilon\vartheta(e_a)})
\ ,$$
 and \eqref{variationgravitino} follows. Finally,  \eqref{variationgravitino1}  follows immediately from 
$$ \d_\varepsilon \psi^b_a=   \left.E^b\left([\varepsilon,  \pi^\cD( e_a)]\right)\right|_{M_o} \!\!\!\! - 
\left.E^b\left(\pi^\cD([\varepsilon,  e_a])\right)\right|_{M_o} \!\!\!\!= \left.E^b(\cL^{g}_{{\varepsilon\ \pi^{\cD}(e_a)}}) \right|_{M_o}\ .
\eqno\square$$
\end{pfns}

\begin{prop} \label{transformations2} Given a   spinor field $\varepsilon = \varepsilon^\a E_\a|_{M_o}$,  the components of the corresponding (super) variation of the A-field  are  of the form
 \beq\label{transA} \d_\varepsilon \Gamd \a a \b =  E^\d(\bD_{e_a}\ve)  \Gamb \a \d \b + \ve ^\g \left(\Gamd \d a \b \Gamb \a \g \d + \Gamd \d a \g  \Gamb \a \b \d  -  R_{\g a \b}^{\phantom{\g a} \a} - R_{a \b \g }^{\phantom{a \b} \a} - \psi^\d_a R_{\b \g \d}^{\phantom{\b \g} \a} \right) + $$
 $$ + \ve^\g\left(   \Gamd \z a \b \Gamb \a \z \g  -
  \psi_a^\z\psi_c^\xi \cL_{\z\beta}^c \Gamb \a \xi \gamma + 
  \psi_a^\z \cL^c_{ \z \b} \Gamd \a c \g + \Gamc \a a \g \b  - \Gamb \a \b {\g|a} \right) \eeq
  \beq\label{transAbis} \d_\varepsilon \Gamd b a \b =  \ve^\g \cL^b_{\g \d} \Gamd \d a \b\ ,\eeq
where:
\begin{itemize}
\item[--] where $\Gamb \a \g \b$, $\Gamc \a a \g \b$, $R_{A B C}^{\phantom{A B} D}$ are the components of B- and C-fields  and of the Riemann tensor $R$ of $\n$   w.r.t. $(E_a|_{M_o}, E_\a|_{M_o}, e^a ,E^\alpha|_{\cS})$; 
\item[--] $\Gamb \a \b {\g|a} \= E^\a\left((\bD_{e_a} \bB)_{E_\b E_\g} - \left(\bA_{e_a} \cdot \bB\right)_{E_\b E_\g}\right)$, where $\bA_{e_a} \cdot$ denotes the natural action of 
$\bA_{e_a} |_x \in \Hom(\cD_x, \cD_x)$ on $\bB_x \in \Hom (\cD_x \times \cD_x, \cD_x)$. 
\end{itemize}
\par
\end{prop}
\begin{pf}  As in the previous proof, for simplicity of notation, we  denote by ``$\varepsilon$'' also  the extension $X^{(\varepsilon)} \in \gX_{loc}(M,\cS)$. By definition of Lie derivative and   first Bianchi identity, we have that 
$$\left(\cL_\varepsilon (\pi^\cD \circ T)\right)_{YZ} =\left[\varepsilon, (\pi^\cD \circ T)_{Y Z} \right] -  (\pi^\cD \circ T)_{[\varepsilon, Y] Z} 
 -  (\pi^\cD\circ T)_{Y [\varepsilon, Z]} = $$
 $$ = (\cL_\varepsilon -\n_\varepsilon)(\pi^\cD ( T_{Y Z})) +$$
 $$+  \pi^\cD\left(  (\n_\ve T)_{YZ} + T_{\n_\ve Y Z} + T_{Y \n_\ve Z} -  T_{[\varepsilon, Y] Z} - T_{Y [\varepsilon, Z]}\right) = $$
 $$ =(\cL_\ve-\n_\ve)(\pi^\cD (T_{Y Z})) + $$
 $$ + \pi^\cD\left(  (\n_\ve T)_{YZ} + T_{\n_Y \ve Z} + T_{Y \n_Z \ve} +  T_{T_{\ve Y} Z}+ T_{ T_{Z \ve} Y} \right)  = $$
 $$ \overset{\text{Bianchi\ id.}}= (\cL_\ve -\n_\ve)(\pi^\cD (T_{Y Z})) + \pi^\cD \left(T_{\n_Y \ve Z} +   T_{Y \n_Z \ve } \right)+$$
 \beq \label{Bianchi} +   \pi^\cD \left(R_{\ve Y} Z + R_{YZ} \ve +   R_{Z \ve} Y-T_{T_{YZ} \ve} -  (\n_Y  T)_{Z\ve}- (\n_Z  T)_{\ve Y} \right)\ .\eeq
On the other hand,  
 $$\d_\varepsilon \Gamd \a a \b = E^\a((\d_\ve \bA)_{e_a E_\b}) =  - \left.E^\a(\left(\cL_\varepsilon (\pi^\cD \circ T)\right)_{e_a E_\b})\right|_{M_o}\ .$$
 Hence, from \eqref{Bianchi}, we get that
 $$\d_\varepsilon \Gamd \a a \b =  - E^\d (T_{e_a E_\b})E^\a([\ve,  E_\d ] - \n_\ve E_\d )   -  E^\a\left(T_{\n_{e_a} \ve E_\b}  \right) -  $$
 \beq \label{Bianchi1} -   E^\a \left(R_{\ve e_a} E_\b + R_{e_a E_\b} \ve +   R_{E_\b \ve} e_a -T_{T_{e_a E_\b} \ve} -  ( \n_{e_a}  T)_{E_\b \ve}- (\n_{E_\b}  T)_{\ve e_a} \right)= $$
 $$ =    \ve ^\g \Gamd \d a \b \Gamb \a \g \d + E^\d(\n_{e_a}\ve) \Gamb \a \d \b  - \ve^\g\left(R_{\g a \b}^{\phantom{\g a} \a} + R_{a \b \g }^{\phantom{a \b} \a} + \psi^\d_a R_{\b \g \d}^{\phantom{\b \g} \a} \right)+$$
 $$ +\ve^\g E^\a\left(T_{T_{e_a E_\b} E_\g} +
(\n_{e_a}  T)_{E_\b E_\g} + (\n_{E_\b}  T)_{E_\g e_a} \right) \ . \eeq
Now,
%using the definitions of  A-field and B-field and the properties of being strict Levi-Civita
we  remark  that at the points of $M_o$, 
 \begin{itemize}
\item[1)] $E^\d(\n_{e_a}\ve) = E^\d(\bD_{e_a}\ve) - \varepsilon^\g\Gamd \d a \g $; 
\item[2)] $E^\a(T_{T_{e_a E_\b} E_\g})   =  \Gamd \z a \b \Gamb \a \z \g  + \psi_a^\z \cL^c_{ \z \b} \Gamd \a c \g-
\psi_c^\xi\psi_a^\z \cL^c_{ \z \b} \Gamb \a \xi \g$.
%\item[(ii)] $E^\a(T_{T_{e_a E_\b} E_\g})   =  \Gamd \z a \b \Gamb \a \z \g  + 
%\psi_a^\z \Gamb \e \z \b \Gamb \a \e \g + 
%\psi_a^\z \cL^c_{ \z \b} \Gamd \a c \g$.
 \end{itemize}
 Replacing (1) and (2)  in \eqref{Bianchi1},  we get \eqref{transA}.
 Similarly, from \eqref{Bianchi}, 
 $$\d_\varepsilon \Gamd b a \b =  - \left.E^b\left(\left[\varepsilon,  \pi^\cD (T_{e_a E_\b})\right]\right)\right|_{M_o}\ ,$$
 from which \eqref{transAbis} follows immediately.
% 
% 
%Setting $Y = \wh e_i$, $Z = E_\a$ and $X\in \G(\cD)$ as in (\ref{spinorone}), we get that on $M_o$
% $$ (\d_X \bA_{i \a}^\b) E_\b \simeq -R_{X\wh e_i} \cdot E_\a -   R_{\wh e_i E_\a} \cdot X - R_{E_\a X}\cdot \vartheta(\wh e_i)  + \bA_{\cL^g(\vartheta(\wh e_i), E_\a) X}   -  \bB_{\bA_{\wh e_i E_\a} X} + $$
%$$+ (\wt \n_{\wh e_i} \bB)_{E_\a X}  - (\wt \n_{E_\a} \bA)_{ X\wh e_i }  
% - \bB_{\wt \n_{\wh e_i}   X  E_\a} + 
%  \bA_{\wh e_i \wt \n_{E_\a} X}-\bB_{X\bA_{\wh e_i E_\a}}  =$$
% $$=-R_{X\wh e_i} \cdot E_\a -   R_{\wh e_i E_\a} \cdot X - R_{E_\a X}\cdot \vartheta(\wh e_i)  + \bA_{\cL^g(\vartheta(\wh e_i), E_\a) X}  + (\wt \n_{E_\a} \bA)_{\wh e_i X } $$
%\beq\label{ancora}+ (\wt \n_{\wh e_i} \bB)_{E_\a X} 
% - \bB_{\wt \n_{\wh e_i}   X  E_\a}\ .
%\eeq
\end{pf}
\begin{cor} \label{corollarytrans} If $\cG$ is strict Levi-Civita and satisfies the constraint
\beq T^\cD \equiv 0\ ,\eeq
then  \eqref{variationgravitino} simplifies into 
\beq
\label{transgrav}
\d_\varepsilon \psi^\a_a=   e_a( \varepsilon^\a) + \varepsilon^\b (\GamD \a a \b + \Gamd \a a \b) \ , 
\eeq
while \eqref{transA} simplifies into
 \beq\label{transA1} \d_\varepsilon \Gamd \a a \b =   \ve ^\g \left( -  R_{\g a \b}^{\phantom{\g a} \a} - R_{a \b \g }^{\phantom{a \b} \a} - \psi^\d_a R_{\b \g \d}^{\phantom{\b \g} \a} 
+  \psi_a^\z \cL^c_{ \z \b} \Gamd \a c \g + \Gamc \a a \g \b  \right)\ . \eeq
 \end{cor}
\section{Classical supergravities as supergravities of type $\gg$}
\label{representing}
In this section, we want to indicate how simple supergravity in four 
dimension
might be encoded in the language of supergravities of type $\gg$. We also
give  short remarks on other supergravities in four and higher dimensions,
supporting the expectation that they  can  be
presented as supergravities of type $\gg$ too.
In the following, the  discussion is forced to be informal. Indeed, a
rigorous presentation of supergravity should be based on various notions 
of supergeometry,
which will be introduced in \cite{SaS}.
%The purpose of this section is to show that  simple 4-dimensional supergravity 
%and other supergravity theories can be presented in the language of  supergravities of type $\gg$.\par
\subsection{Notations} In all the following,      $\cG = ((M, M_o, \cD), (g,  \n) )$ is  a fixed (super) gravity of type $\gg$. 
\subsubsection{Clifford product between elements of $TM$ and  $\cD$}
For any $x \in M$,  $w \in T_xM$ and $\imath^{(x)} \in  O_g(M, \cD)$, we denote by $w = w^V + w^S$ the $g$-orthogonal decomposition of $w$ into
$\cD^\perp$- and $\cD$- components and we set 
$$\wh w = \imath^{(x)-1} (w) \in V + S\ ,\ \ 
\wh w^V = \imath^{(x)-1} (w^V) \in V \ , \ \ \wh w^S = \imath^{(x)-1} (w^S) \in S\ .$$
%where we use $(\cdot)^V$ and $(\cdot)^S$  to denote  $V$- and $S$-components, respectively. Notice that
% $\wh w^V$, $\wh w^S$ and $\wh s$ are uniquely determined by $w$ and $s$, up to an element the structure group of $\operatorname{O}_g(M, \cD)$. \par
 For any  $s  \in \cD_x$, we call 
  {\it Clifford product between $w$ and $s$} the element in $\cD_x$
\beq \label{cliff1} w \cdot s \= \imath^{(x)}(\wh w^V \cdot \wh s) \ , \eeq
where ``$\wh w^V \cdot \wh s$''  is the  usual Clifford product.  One can check that \eqref{cliff1} does not depend on the choice of 
$\imath^{(x)} \in  O_g(M, \cD)$.  We  extend canonically \eqref{cliff1} to a product $\a \cdot s \in \cD_x$ between any $\a \in \L T_x M$ and  $s \in \cD_x$ and, by $g$-duality,  also to a product $\o\cdot s$ between any $\o \in \L T^*_x M$ and $s \in \cD_x$. \par
\smallskip
We remark that {\it any  such Clifford product is preserved by the action \eqref{actionofdiffeomorphisms}\/}.
\subsubsection{$\cD^\perp$-curvatures and Rarita-Schwinger form}
We denote by $\Ricperp$  and $s^{\cD^\perp}$ the tensor field  and the scalar function,  defined  at any $x \in M$ by  
$$ \Ricperp_x(v_1, v_2) = \sum_{i = 1}^{n} \e_i g((\pi^{\cD^\perp} \circ R)_{v_1 E_i} v_2,  E_i), \ \  s^{\cD^\perp}=\sum_{j = 1}^n\epsilon_j\Ricperp (E_j,E_j)\ ,$$
where $(E_i)$ is any $g$-orthonormal basis of $\cD^\perp_x$ and $\e_i = g(E_i, E_i) = \pm 1$. These objects are related 
with Ricci and scalar curvature of the metric connection $D$ on $M_o$ as follows.  Since the curvature $R^D$ of $D$ is 
at any $x \in M_o$ given by 
$$ R^D_x = ( \pi^{\cD^\perp}|_{TM_o})^{-1}\left(\pi^{\cD^\perp} \circ R|_{TM_o \times TM_o \times TM_o}\right)\ ,$$
we get that Ricci curvature $\operatorname{Ric}^D$ and scalar curvature $s^D$ of $D$ are given by
\beq \label{ricci1} \operatorname{Ric}^D_x(v_1, v_2) = \Ricperp(v_1, v_2) + \sum_{i = 1}^{n}\e_i g ((\pi^{\cD^\perp}\circ R)_{v_1 \pi^{\cD}(e_i) }v_2, e_i)\ , \eeq
\beq \label{ricci2}
s^{D}_x=s^{\cD^\perp}_x+\sum_{i,j= 1}^n\e_i\e_jg_x((\pi^{\cD^\perp}\circ R)(\pi^{\cD}(e_j),\pi^{\cD}(e_i))e_j,e_i)\ , 
\eeq
for any $v_1, v_2 \in T_x M_o$, where $(e_i)$ is a $\wh g$-orthonormal basis for $T_{x}ÊM_o$. \par
%
%
%
%\beq Ric^D(v_1, v_2) = \sum_{1}^{n}\e_i\wh g (R^{D}_{v_1 e_i}v_2,e_i)=\sum_{1}^{n}\e_i g (\pi^{\cD^\perp}\circ R_{v_1 e_i}v_2,E_i) = $$
%$$ =\sum_{1}^{n}\e_i g ((\pi^{\cD^\perp}\circ R)_{v_1 E_i}v_2,E_i) + \sum_{1}^{n}\e_i g ((\pi^{\cD^\perp}\circ R)_{v_1 E_\a }v_2,E_i)  E^\a(e_i)= $$
%$$ = \sum_{1}^{n}\e_i g ((\pi^{\cD^\perp}\circ R)_{v_1 E_i}v_2,E_i) + \sum_{1}^{n}\e_i g ((\pi^{\cD^\perp}\circ R)_{v_1 \pi^\cD(\cdot) }v_2, \pi^{\cD^\perp}(\cdot )) (e_i, E_i)\eeq
%$$\sum_{1}^{n}\e_i g ((\pi^{\cD^\perp}\circ R)_{v_1 \pi^\cD(\cdot) }v_2, \pi^{\cD^\perp}(\cdot )) \in \cD^* \otimes \cD^\perp{}^*  --- (V + S) \otimes (V+ S) \otimes V \otimes S$$
%where $(e_i)$ is any $g$-orthonormal basis of $T_x M$. 
%\sideremark{Check this!}
%Next lemma, whose proof is left to the reader,  shows that these tensors are actually related with  the metric connection $D$. \par
%
%We conclude with the following definition. \par
\medskip
 We call  {\it Rarita-Schwinger 3-form}  the tensor field $\cR \in \L^3 T^*M \otimes_M \cD$ defined at any $x \in M$, $v_1, v_2, v_3 \in T_x M$, by 
$$ \cR_x(v_1, v_2, v_3) \= \sum_{\s \in P_3} (-1)^{\e(\s)}v_{\s(1)} \cdot  \left((\pi^\cD \circ T)_x\left(v_{\s(2)}, v_{\s(3)}\right)\right)\ .$$
%where $\cH^{\L^2 \cD^\perp; \cD}_x = - \cL^g_x$ is the torsion component  described in \eqref{decoT}. \par
Using coordinates on $M_o$, the 3-form $\left.\cR\right|_{\L^3 TM_o}$ is related with  $\vartheta$ by
\be \label{RS} \left(\cR|_{\L^3 TM_o}\right)_{\frac{\partial}{\partial x^{i_1}}, \frac{\partial}{\partial x^{i_2}}, \frac{\partial}{\partial x^{i_3}}}= 2
\sum_{\s \in P_3}ÿ(-1)^{\e(\s)} \left(\frac{\partial}{\partial x^{i_{\s(1)}}}\cdot  \n_{\frac{\partial}{\partial x^{i_{\s(2)}}}}
\left(\vartheta(\frac{\partial}{\partial x^{i_{\s(3)}}})\right)\right)\, .
\ee
%Given a smooth orientation  for the spaces $\cD^\perp_x$,  $x \in M$, we denote the corresponding Hodge-star operator, determined by  $g|_{\cD^\perp_x \times \cD^\perp_x}$, by the symbol
%$$\star_x: \Lambda^p (\cD^\perp_x)^* \longrightarrow  \Lambda^{n-p} (\cD^\perp_x)^*\ .$$
%\par
%We call {\it Rarita-Schwinger $n-2$-form} the tensor field 
%$\star\cR \in \Lambda^{n-2} T^* M\otimes_M \cD$  defined by 
%\beq \star \cR_x(\cD_x, \cdot, \dots, \cdot) = 0\ ,\qquad  \star \cR_x|_{\cD^\perp_x \times \dots \times \cD^\perp} = \star_x (\pi^{\cD}\circ T)|_{\cD^\perp_x \times \cD^\perp_x}\ .\eeq 
%\par
%\bigskip
%If $n = 4$ and the signature of $g|_{\cD^\perp_x \times \cD^\perp_x}$ is $(p,q)=(3,1)$, one has that 
%$$\star \cR(\cdot,\cdot)= -\frac{1}{4}
%\varepsilon^{j_1 j_2 j_3 j_4} (\pi^{\cD}\circ T)(E_{j_1},E_{j_2}) g(E_{j_3},\cdot) g(E_{j_4},\cdot) 
%$$
%where $(E_1, \dots, E_4)$ is a positively oriented $g$-orthonormal frame field for $\cD^\perp$. \par
\subsection{Simple 4D-supergravity} \label{section52} Let $V = \bR^{3,1}$ and $\gg = \so(V) + V + S$ the super-Poincar\`e algebra determined by the admissible bilinear form 
$\b(s,s') =   \Re\o(s,s')  = - \Re(is^T \G_0\G_2 s')$ on the irreducible spinor module  $S = S^+ + S^- $ of $\cC \ell_{3,1}$
 (see  Example \ref{4Dexample}).  \par   %\begin{example}
\par
\bigskip
Simple $4D$-supergravity can be interpreted as  a supergravity 
$$\cG = ((M, M_o, \cD = \cD^+ + \cD^-), (g , \n))
$$
of type $\gg$ (\footnote{For this super-algebra,  the space
$\gh$ is trivial (see Remark \ref{Ctriviality}) and 
  $ \n = \n^o$ for $\cG$.}), subjected to the following constraints 
 and equations, which are equivalent to Wess and Zumino's constraints  and the usual Euler-Lagrange equations 
  (\cite{WZ, WB, vN}). \par
  \bigskip
\centerline{\bf Constraints}
\vskip0.2cm
  \begin{itemize}
\item[1)]  $\n$  is strict Levi-Civita (i.e. $T^{\cD^\perp} = 0 = \cC^{\cD, \cD^\perp; \cD^\perp}$);  
\item[2)]  $T^\cD \equiv 0$. 
\end{itemize}
\vskip0.3cm
\centerline{\bf Equations}
\vskip0.2cm
\begin{itemize}
\item[i)] $ \left.\cC^{\cD, \cD^\perp; \cD}\right|_{\cS \otimes TM_o} = 0$\,\,\, ({\it vanishing of auxiliary fields}); 
\vskip0.2cm
\item[ii)] $ \cR|_{\L^3 T M_o} = 0$\,\,\, ({\it Rarita-Schwinger eq.}); 
\vskip0.2cm
\item[iii)] $
\left.\Ricperp\right|_{T M_o \times TM_o}\!\!\!\!\!-\frac{1}{2}\left. s^{\cD^\perp}g(\pi^{\cD^\perp}(\cdot),\pi^{\cD^\perp}(\cdot))\right|_{T M_o \times TM_o}\!\!\!\!\!\!=0$\,\,\,
({\it Einstein eq.}). 
\end{itemize}
\par
\medskip
%$$ \left.\Ricperp\right|_{T M_o \times TM_o}   - 
%$$
%\k\left. \sum_{i = 1}^{n + m} 
%g\left(\cdot, \cL^g( \pi^{\cD}(e_i), \G_5(\star \cR(e_i, \cdot)))\right) \right|_{T M_o \times TM_o} = 0\ , 
%$$
%$$
%for some suitable constant $\k$. 
The first equation corresponds to the vanishing of the  ``auxiliary fields'', the 
second one to the so-called Rarita-Schwinger equation for gravitinos while the last one corresponds to the Euler-Lagrange equation for gravitons. \par
\medskip
Firstly, from constraints (1) and (2) and Bianchi identities, one gets that the A-field $\bA$
 is of the following very special form (see \cite{WB}, Ch. XV) 
\beq  \label{5.5} \bA_{X}s   =\left.\cC^{\cD, \cD^\perp; \cD}\right|_{\cS \otimes TM_o} (s,X)= 
$$
$$=- \Re(a)X\cdot \G_5\cdot s +  i \Im(a)  X \cdot  s + iA(X)\G_5\cdot s + \frac{i}{3}X\cdot A\cdot\G_5\cdot s
%\cA(X)s
 %(the operator \G_5 is as in \eqref{gammafive})
\eeq
for a complex function $a: M_o \rightarrow \bC$ and a 1-form $A \in T^* M_o$, usually called {\it auxiliary fields}, and hence that  
(i) %$\left.\cC^{\cD, \cD^\perp; \cD}\right|_{\cS \otimes TM_o} = 0$
is equivalent to equations $a = 0$, $A = 0$.
\par
\medskip
Equation (ii) is equivalent to the Rarita-Schwinger equation  by simply comparing the coordinate expression  \eqref{RS} with  \cite{vN}, formula (5) at p. 222. \par
 %(denoted by    ``$M$'' and ``$(b_a)$''  in \cite{WB}).\par 
\medskip
Now, assume constraints (1), (2) and equations (i), (ii) hold. By equations \eqref{ricci1}, \eqref{ricci2} and Bianchi identities, one can prove $s^{\cD^\perp} = s^D$ so that (iii) reads
\beq\label{einstein1} \left( \operatorname{Ric}^D(X, Y)   - \frac{1}{2} s^{D}\wh g(X,Y) \right) - \sum_{i = 1}^{n}\e_i g ((\pi^{\cD^\perp}\circ R)_{X \pi^{\cD}(e_i) }Y, e_i) = 0 \eeq
for any $X,Y\in \gX(M_o)$.
Using again Bianchi identities and (ii), 
the equation \eqref{einstein1} becomes equivalent to 
%\beq \left( \operatorname{Ric}^D  - \frac{1}{2} s^{D} \right) - \sum_{i = 1}^{n}\e_i g ((\pi^{\cD^\perp}\circ R)_{(\cdot) \pi^{\cD}(e_i) }(\cdot), e_i) = 0\ ,\eeq
$$ \left( \operatorname{Ric}^D(X,Y)  - \frac{1}{2} s^{D}\wh g(X,Y) \right) - \sum_{i = 1}^{n}\e_i g (
\pi^{\cD}(e_i), X \cdot \left( \pi^{\cD} \circ T)_{Y e_i}\right) = 0\ ,
$$
for any $X,Y\in \gX(M_o)$.
From the expression in  coordinates 
$$ \pi^{\cD} \left( T_{\frac{\partial}{\partial x^i} \frac{\partial}{\partial x^j}}\right) = \n_{\frac{\partial}{\partial x^{i}}}
\left(\vartheta(\frac{\partial}{\partial x^{j}})\right) - \n_{\frac{\partial}{\partial x^{j}}}
\left(\vartheta(\frac{\partial}{\partial x^{i}})\right)$$
and \cite{vN}, formula (10) at p. 222, one gets that (iii) is equivalent to the usual Euler-Lagrange equations for gravitons  (see \cite{vN}, formula (6) at p.222). 
  \par
%and 
%$$
%\cA(X)\in\End(S^\pm)\ ,
%$$
%$$
%\cA(r\otimes t)(s^++s^-)=-i\sum_{i<j}A(\G_{ij}r,t)\G^{ij}s^+\  +\ i\sum_{i<j}A(r,\G_{ij}t)\G^{ij}s^-\ .
%$$ 
%Up to factors, {\it  $A$ and  $a$ coincide with   the  fields  denoted  ``$(b_a)$'' and  ``$M$''  in  \cite{WB} and  ``$(A_a)$''  and ``$P + i S$'' in  \cite{vN}, 
%i.e. the {\rm auxiliary fields} of simple $4D$-supergravity}. Moreover $\left.\cC^{\cD, \cD^\perp; \cD}\right|_{M_o} = 0$
%is equivalent to  $A = 0$, $a = 0$. \par
%
\medskip
Finally, we remark that, under the constraints (1) and (2),  the usual  transformation rules for  graviton and gravitino (see \cite{WB}, Ch. XVIII) coincide with those in Proposition \ref{transformations1} and Corollary \ref{corollarytrans} and it is   reasonable to expect that, via \eqref{5.5},  the usual transformation rules of auxiliary fields imply  the variations for the A-field, determined in Corollary \ref{corollarytrans}. We plan to check carefully this point in the near future. \par
\medskip
In any case, we claim  that  the above constraints and equations are {\it manifestly covariant} and 
hence {\it invariant under all super-variations\/} of Definition 4.4, %\ref{spinorone}, 
by the following reasons.    \par
\smallskip 
Consider
the system  $\cE$  on  $(\cD, g, \n)$  given by the tensorial equations
$$T^{\cD^\perp} =0\ , \quad  \cC^{\cD, \cD^\perp; \cD^\perp} = 0 \ , \quad T^\cD = 0\ ,$$
$$\cC^{\cD, \cD^\perp; \cD}= 0\ ,\quad   \cR = 0 \ , \quad \Ricperp -\frac{1}{2} s^{\cD^\perp}g(\pi^{\cD^\perp}(\cdot),\pi^{\cD^\perp}(\cdot))=0 \ . $$
Any (local) solution of $\cE$  gives physical fields  satisfying  the system $\cE_o$ of  (1), (2),  (i), (ii), (iii).  So, being $\cE$ of tensorial type, 
 in order to check the  manifest covariance, it remains  to show  that  any (local) solution of $\cE_o$  is given by the physical  fields of  some 
 (local) solution of $\cE$.   
 \par
 Indeed, following  the same   arguments   used in  \cite{SaS}  to check  the manifest covariance of the 11D supergravity equations and constraints,  one can   see  that the  conditions  $\cC^{\cD, \cD^\perp; \cD^\perp} = T^\cD = \cC^{\cD, \cD^\perp; \cD}= 0 $, together with the relations $R|_{\cD\otimes\cD}=0$ and (15.21) of  \cite{WB} (which come from  the  first Bianchi identities of $\nabla$),  coincide  with  the    {\it  rheonomic constraints}  considered by   Castellani, D'Auria and Fr\`e  in  \cite{CDF}, Ch.III.3.5. By the  results of \cite{CDF}, one gets that  all equations of  the system  $\cE$  are consequences  of such rheonomic constraints and   Bianchi identities and   that  the required one-to-one correspondence between  solutions of $\cE_o$  and  solutions of $\cE$  is  a corollary of  the properties of the  rheonomic constraints (we refer to \cite{SaS}  for more details on this line of arguments).
%A similar question is faced and solved with  positive answer 
 %in the so-called {\it rheonomic approach\/} to $4D$-supergravity (\cite{CDF}, Ch. III.3). We expect that analogous %arguments, based on a variational origin 
%of the equations,   can  be used in our setting. We will  discuss this and related questions in our future %investigations. 
\subsection{Other supergravities}
\subsubsection{Gates and Siegel's supergravities}
Simple 4D-supergravity is one  of the supergravities, parameterized by   $\z \in \bR \cup \{\infty\}$, introduced by Gates, Siegel in \cite{SG, GS} (see also \cite{RS, RS1}).  
All of them can be interpreted  as   supergravities 
$$\cG = ((M, M_o, \cD= \cD^+ + \cD^-), (g, \n))$$ 
of the same type $\gg$ of simple supergravity and they are subjected to the following constraints for $\z \neq -\frac{1}{3}$ (the case  $\z = -\frac{1}{3}$ is simple supergravity).
\vskip0.3cm 
\centerline{\bf Constraints}
\vskip0.2cm
  \begin{itemize}
\item[1)]  $\n$  is (non-strict) Levi-Civita  with  $\cC^{\cD, \cD^\perp; \cD^\perp}_x$ of the form
$$ \cC^{\cD, \cD^\perp; \cD^\perp}_x = \frac{\z}{3 \z + 1} (\Re(\cT) \circ \pi^{\cD}-i\Im(\cT) \circ \pi^{\cD}\circ\G_5)_x \otimes \pi^{\cD^\perp}_x\ ,
$$
%$$ \cC^{\cD, \cD^\perp; \cD^\perp}_x = \frac{\z}{3 \z + 1} (\cT \circ \pi^{\cD^+})_x \otimes \pi^{\cD^\perp}_x+ \frac{\z}{3 \z + 1} (\overline{\cT} \circ \pi^{\cD^-})_x \otimes \pi^{\cD^\perp}_x
%$$
for some complex-valued 1-form $\cT \in T^* M$;  
\item[2)] $T^{\cD}$ is of the form 
$$ T^\cD_x(v_1, v_2)  =  \frac{1}{2}ÿ\frac{\z + 1}{3 \z +1}\left(\pi^{\cD^\pm}_x(v_1) (\Re(\cT) \circ \pi^{\cD^\pm}-i\Im(\cT)\circ\pi^{\cD^\pm}\circ\G_5)_x(v_2)\right.+
$$
$$
\phantom{cccccccccccccccccccc}+\left.\pi^{\cD^\pm}_x(v_2) (\Re(\cT) \circ \pi^{\cD^\pm}-i\Im(\cT)\circ\pi^{\cD^\pm}\circ\G_5)_x(v_1)
\right)+
$$
$$
 \phantom{ccccccccccc}+\frac{1}{2}ÿ\frac{\z - 1}{3 \z +1}\left(\pi^{\cD^\mp}_x(v_1) (\Re(\cT) \circ \pi^{\cD^\pm}-i\Im(\cT) \circ \pi^{\cD^\pm}\circ\G_5)_x(v_2)\right.+
$$
$$
\phantom{ccccccccccccccccccc}+\left.\pi^{\cD^\mp}_x(v_2) (\Re(\cT) \circ \pi^{\cD^\pm}-i\Im(\cT) \circ \pi^{\cD^\pm}\circ\G_5)_x(v_1)\right)
$$
for any $v_1$, $v_2$ $\in T_x M$; 
%\beq T^\cD_x(v_1, v_2)  =  \frac{1}{2}ÿ\frac{\z + 1}{3 \z +1}\left( \pi^{\cD^+}_x(v_1) (\cT \circ \pi^{\cD^+})_x(v_2) +  \pi^{\cD^+}_x(v_2) (\cT \circ \pi^{\cD^+})_x(v_1)\right)+
%$$
%$$
%+\frac{1}{2}ÿ\frac{\z + 1}{3 \z +1}\left( \pi^{\cD^-}_x(v_1) (\overline{\cT} \circ \pi^{\cD^-})_x(v_2) +  \pi^{\cD^-}_x(v_2) (\overline{\cT} \circ \pi^{\cD^-})_x(v_1)\right)+
%$$
%$$
%\frac{1}{2}ÿ\frac{\z - 1}{3 \z +1}\left( \pi^{\cD^+}_x(v_1) (\overline{\cT} \circ \pi^{\cD^-})_x(v_2) +  \pi^{\cD^-}_x(v_2) (\cT \circ \pi^{\cD^+})_x(v_1)\right)
%$$
%$$
%\frac{1}{2}ÿ\frac{\z - 1}{3 \z +1}\left( \pi^{\cD^+}_x(v_2) (\overline{\cT} \circ \pi^{\cD^-})_x(v_1) +  \pi^{\cD^-}_x(v_1) (\cT \circ \pi^{\cD^+})_x(v_2)\right)
%\eeq 
%for any $v_1$, $v_2$ $\in T_x M$; 
\item[3)] the torsion  components  $\cC^{\cD^+,\cD^\perp ; \cD^-}ÿ$ and $ \cC^{\cD^-, \cD^\perp; \cD^+} $
%, 
%taking values in $\Hom(\cD_x^+ \times \cD_x^\perp, \cD^-_x)$ 
%and $\Hom(\cD_x^- \times \cD^\perp_x, \cD^+_x)$, 
vanish. 
%(note: this constraint is redundant in our setting, since  we assume that $\wt \n$ preserves the distributions $\cD^{\pm}$). 
\end{itemize}
%\vskip 1 cm
These constraints are manifestly covariant.  We expect that also  the Euler-Lagrangian  equations of these supergravities 
are manifestly covariant, as it occurs for  simple 4D supergravity.  
%
%As before, these constraints are manifestly covariant. It is reasonable to expect that also  the Euler-Lagrangian  equations of these supergravities 
%are manifestly covariant, as  it occurs for  simple 4D supergravity.  \par
%\smallskip
%
%We recall that the spinor module $S$ (see Example \ref{4Dexample}) of simple $4D$-supergravity is usually endowed with an anti-linear involution 
%\be
%\label{involuzione}
%s\longrightarrow s^c:=-iC\G_0 \overline{s}\ 
%\ee
%which commutes with the action of $\Cl^{0}_{3,1}$. The fixed point set of the conjugation \eqref{involuzione} is the so-called space $\bM$ of pseudo-Majorana spinors. Due to this, it is customary to consider $S$ endowed with the admissible complex-valued bilinear form 
%\eqref{volumeform}
%and $\gg = \so(V) + V^\bC + S$ to be the associated super Poincar\`e algebra.
%In our setting, the real super Poincar\'e algebra to be considered would be simply $\so(V)+V+\bM$ with bracket $[\bM,\bM]\subset V$ induced by the form $\beta_1= \Re\o$. 
%\par
%For simplicity, we denote by $(M, M_o, \cD = \cD^+ + \cD^-)$ a space-time of type $\gg$, having in mind that all geometric objects should be compatible with the conjugation \eqref{involuzione}. Finally, we will denote by $\G_5$ the tensor field on $\cD$ of type $(1,1)$ which
%acts on each space $\cD_x \simeq S$ as the gamma matrix \eqref{gammafive}, i.e.
%$$ \G_5 = - \pi^{\cD^+} + \pi^{\cD^-}\ ,$$
%where $\pi^{\cD^\pm}$ denote the $g$-orthogonal projections on $\cD^\pm$.
%\subsection{11D-supergravity}\hfill\par
\subsubsection{Supergravities in dimensions $n \geq 5$}  We recall that
the Poincar\`e superalgebra $\gg = \so_{3,1} + \bR^{3,1} + S$ of simple 4D supergravity is  the algebra of rigid supersymmetries of 
 maximally supersymmetric vacua solutions and that the theory is actually determined  by ``gauging''  such
symmetries. \par
\smallskip
Supergravities   in dimensions $n \geq 5$  are  similarly obtained from algebras $\gg$ of  rigid supersymmetries  
of  homogenous manifolds  playing the role of   vacua. \par
\smallskip
The  superalgebra $\gg=\gg_0+\gg_1$  is usually taken from Nahm's classification (\cite{Na}), i.e. it is a simple Lie superalgebra with   $\gg_0 = \gp \oplus\gk$,  where  $\gk$  is reductive and  $\gp$  is a conformal or de Sitter algebra,  and with  $\gg_1 = S$ a spinor module.
\par
The associated  simply connected,   homogeneous supermanifold is of the form $G/H$, with  $ \gh = \so_{p,1}  \oplus \gk \subset
 \gp \oplus \gk$ and it is endowed with the
  $G$-invariant distribution $\cD$  with  $\cD|_{eH} = S$. Its
 Levi form at $ e H$  is the  $\so_{p,1}$-invariant  tensor 
$$L \in S^2 S^* \otimes  \bR^{p,1}\ ,\qquad L(s,s') = [s, s'] \mod \gh\ .$$
This means that {\it $(G/H, G_0/H, \cD)$ is a space-time 
of type $\gg'$}, where $\gg'$ is  the super Poincar\`e algebra $\gg' = (\so_{p,1} + \bR^{p,1}) + S$,  with brackets $[\cdot, \cdot]|_{S \times S} = L$.\par
% We stress the fact that this occurs for {\it any} superalgebra $\gg$ of Nahm's list. \par
\medskip
\bigskip
\bigskip

\bibliographystyle{my-h-elsevier}

\end{document}